\documentstyle[12pt,aasms4,psfig]{article}

\slugcomment{Accepted for publication in Astrophysical Journal}

\lefthead{Chiappini et al.}
\righthead{The Chemical Evolution of the Galaxy:
the two-infall model}

\begin{document}

\title{The Chemical Evolution of the Galaxy:\\
the two-infall model}

\author{C. Chiappini\altaffilmark{1,2}, F. Matteucci\altaffilmark{1}}

\affil{Dipartimento di Astronomia, Universit\'a di Trieste,
SISSA, Via Beirut 2-4, I-34013 Trieste, Italy}
\affil{Instituto Astron\^omico e Geof\'{\i}sico, Universidade de S\~ao Paulo,
S\~ao Paulo, Brazil}

\and

\author{R. Gratton\altaffilmark{3}}
\affil{Osservatorio Astronomico di Padova, 
Vicolo dell'Osservatorio 5, 35100, 
Padova, Italy}





\begin{abstract}
In this paper we present a new chemical evolution model for the Galaxy
which assumes two main infall episodes 
for the formation of halo-thick disk and thin disk, respectively.
We do not try to take into account
explicitly the evolution of the halo since our model is better
suited for computing the evolution of the disk (thick plus thin)
but we implicitly assume that the timescale for the formation
of the halo was of the same order as the timescale for the 
formation of the thick disk.

The formation of the thin-disk is much longer than
that of the thick disk, implying that the infalling gas forming
the thin-disk comes not only from the thick disk but mainly from
the intergalactic medium.

The timescale for the formation of the thin-disk is assumed to be a
function of the galactocentric distance, leading to an inside-out picture
for the Galaxy building. 
The model takes into account the most
up to date nucleosynthesis prescriptions
and adopts a 
threshold in the star formation process which
naturally produces a hiatus in the
star formation rate at the end of the thick disk phase, 
as suggested by recent
observations.
The model results are compared with
an extended set of observational constraints
both for the solar neighbourhood and the whole disk. 
Among these constraints, the tightest
one is the metallicity distribution of the
G-dwarf stars for which new data are now available.
Our model fits very well these new data.

The model predicts also the evolution of the gas mass, the star formation rate,
the supernova rates and the abundances of 16 chemical elements as functions of time and
galactocentric distance. We show that in order
to reproduce most of these constraints
a timescale
$\le 1$ Gyr for the (halo)-thick-disk  
and of 8 Gyr for the thin-disk formation 
in the solar vicinity are required. 

We predict that the radial abundance gradients in the inner regions of the disk
($R< R_{\odot}$) are steeper than in the outer regions, 
a result confirmed by recent abundance determinations, and that the inner ones
steepen in time during the Galactic lifetime. 
The importance and the advantages
of assuming a threshold gas density for
the onset of the star formation process is discussed. 
\end{abstract}




%

\section{Introduction}

The study of chemical evolution of our Galaxy has been addressed by several
models in recent years. Generally, a good agreement
between model predictions and observed properties of the Galaxy 
is obtained by models which assume that the disk formed by infalling gas, as
suggested in the pioneering work of Chiosi (1980). Very
different models assuming gas accretion onto the Galactic disk were 
constructed such as, for instance, 
viscous models (Lacey and Fall 1985, Sommer-Larsen 
and Yoshii 1989, 1990, Tsujimoto et al. 1995), inhomogeneous models 
(Malinie et al. 1993), detailed chemical evolution models 
(Matteucci and Greggio 1986, Tosi 1988,
Matteucci and Fran\c cois 1989 (MF89), Pagel 1989, 
Matteucci and Fran\c cois 1992,
Carigi 1994, Giovagnoli and Tosi 1995, 
Ferrini et al. 1994, Pardi and Ferrini 1994, Pardi et al. 1995,
Prantzos and Aubert 1995, Timmes et al. 1995) 
and chemodynamical models (Samland and Hensler 1996, Burkert et al. 1992).  
However, as recently pointed out by Prantzos and Aubert (1995), the model predictions
are usually
compared with different sets of observables by the different authors and
they suggested that a {\it minimal set} of observables should be adopted.

It is also of fundamental importance to relax the
instantaneous recycling approximation (IRA)
and to account for the contribution of type Ia supernovae (SNe).
This allows us to make a correct comparison between models
and observational data, especially because the observed metallicity
is usually represented by the iron abundance.\par
Recently, Tosi (1996 and references therein)
compared different models (without dynamics) for the 
chemical evolution of the Galactic disk
where IRA was relaxed.
She showed that, in spite of the fact that these models were based on very
different assumptions, they all could fit a reasonable
number of the solar neighbourhood constraints,
whereas significant differences were present
in the predictions for the radial distributions of the various
quantities as well as in the history of radial abundance gradients.
These facts suggest that new models should be tested on the observed radial
properties of the disk.
With respect to the available observational constraints, the last years
have been of crucial importance. New data are now available 
concerning
the age-metallicity relation, the G-dwarf metallicity distribution,
the relative abundances of $\alpha$-elements and iron 
and the
radial abundance gradients and they
require the construction of new models. 
In particular, in a recent paper Gratton et al. (1996) showed that the 
distribution of the abundances of  $\alpha$-elements to Fe 
for a large homogeneous
sample of stars in the solar neighbourhood 
seems to indicate a short timescale for the
evolution of the halo and thick disk phases and a sudden
decrease in the star formation in the epoch preceding the formation of the
thin disk.
They identified three kinematically distinct populations:
i) a population made of halo, thick-disk 
and perhaps bulge stars originating from
the dissipative collapse of the halo, ii) a population of thin disk stars 
originating from the even more dissipative collapse of the disk, and finally
iii) a population of thick-disk stars the origin of which should be different
from the others, namely
they should have formed 
in satellite galaxies and then accreted by the Galaxy during the star
formation gap.
This last component should contribute negligibly to the total number of stars
but it should represent $\sim 50\%$ of the stars with [Fe/H] $<-1.0$.

An analogous result was already found by Beers and Sommer-Larsen (1995). These
authors have shown that the thick disk population extends to very low 
metallicities. For instance, according to that work about $30 \%$
of metal-deficient stars in their sample, with metallicities
[Fe/H] $\leq$ 1.5, 
have kinematic properties which are typical of a thick disk population.
This metal-weak tail of the thick disk population could have its
origin in a major accretion episode during the Galaxy evolution (Beers and
Sommer-Larsen 1995 and references therein).
This is a very important new information which induces to consider a different picture
of Galaxy formation than those previously adopted.

Previous models, in fact, (MF89 and Matteucci and Fran\c cois, 1992)
were based on only
one episode of infall during which first formed the halo and then the disk or
(Pardi et al. 1995) were assuming that 
halo, thick and thin disk formed simultaneously but at different rates
(pure collapse picture).
These models, however, are difficult to conciliate with the new
results discussed above. 

In this paper we present a new chemical 
evolution model for the Galaxy which assumes two main infall
episodes. The first one is responsible for the formation of the 
population i) namely, the one made of that fraction of the halo 
and thick disk stars which originated from a fast dissipative collapse
such as that suggested by Eggen et al. (1962).
It is beyond the scope of the present work to describe
explicitly
the evolution of the halo as we are using a formalism based on 
surface densities which is suitable for disk models. 
This phase, however, is included implicitly in our thick-disk phase
by assuming that the timescale for the formation of the internal
halo was of the same order than the timescale for the formation of the 
thick disk (Gratton et al. 1996).
The second infall episode
forms the thin disk component with a timescale much longer than
that of the thick disk formation. This model implies that most
of the galactic thin disk, if not all,
was formed out of accreted extragalactic
material. 

This scenario for the Galaxy formation is not only in agreement
with the results of Gratton et al. (1996) and Beers and Sommer-Larsen 
(1995) but also with the works by Wyse and Gilmore (1992) and
Ibata and Gilmore (1995). In fact, these last papers have shown
that the spheroidal (bulge and halo) and disk (thick and thin disks)
components of our Galaxy have substantial different
angular momentum distributions. This fact
strongly suggests that the previously adopted
picture, where the gas shed from the halo was the main
contributor to the thin disk formation, should
be revised (Pagel and Tautvaisiene 1995).

\par
The aim of this paper is to test the two infall chemical evolution
model with respect to the maximum number
of available observational constraints in the Galaxy. 
Relatively to our previous models this one adopts the most recent 
nucleosynthesis prescriptions published by Woosley and Weaver (1995)
for the elements produced in massive stars, and by Dearborn et al. (1995) for
$^{3}He$  and D abundances. The inclusion of a threshold in the gas density
below which the star formation process stops, makes the model more physical
and produces new important effects.

In section 2 we describe the set of observational 
constraints to be compared with the models. 
In section 3 we present the model assumptions, the nucleosynthesis
prescriptions and the adopted input parameters. 
In section 4 we compare the model predictions
with the observed properties. 
Section 5 presents the conclusions.

\section{Observational Constraints}

A good model of chemical evolution of the Galaxy should reproduce a number
of constraints which is greater than the number of free parameters. Therefore,
it is very important to choose a high quality set of observational data 
to be compared with models predictions. Our set of constraints include:

\begin{itemize}

\item
The relative number of thin disk and metal-poor stars
(halo plus part of thick disk stars) in the solar cylinder

\item
Type I and type II supernova rates at the present time

\item
Solar abundances

\item
Present-day gas fraction

\item
Age-Metallicity relation

\item
Present-day infall rate

\item
Metallicity distributions for disk and metal-poor stars

\item
The variation in the relative 
abundances of the most common chemical elements

\item
Radial profiles for the SFR and gas mass density

\item
Radial abundance gradients
\end{itemize}

\subsection{Solar abundances}
The
solar abundances should represent the chemical composition of the 
interstellar medium in the solar neighbourhood at the time of sun formation 
(4.5 Gyrs ago). However, Cunha and Lambert (1992)
showed that the abundance of oxygen in the Orion nebula 
is smaller by a factor of 2 than the
solar one (Anders and Grevesse 1989), at variance with the  increase
of the metal abundances in the Galaxy with time as predicted by
the chemical evolution models. 
Thus, it is not clear whether the solar composition should be
considered as representative of the local ISM 4.5 Gyrs ago, a
possibility being that the sun was born in a region closer to the 
Galactic center and then moved to the present region.
It should also be noted that the elemental abundances are uncertain.
As discussed by Timmes et al. (1995), given the uncertainties involved,
abundance values falling within a factor of two inside the observed ones
can be considered as being in agreement with the solar data.

\subsection{Age-metallicity relation}
For the age-metallicity relation in the solar neighbourhood 
the most recent data were derived by Edvardsson et al. (1993)
from F stars with $7.7 < R < 9.3$, where R is the galactocentric
distance corresponding to the star birth places. 
These authors present a spectroscopically calibrated age-metallicity 
relationship which constitutes the most accurate data set available
at present. Previous data were based on photometric surveys (Twarog
1980, Carlberg et al. 1985, Meusinger et al. 1991)
which did not contain
kinematic information. However, this more accurate
age-metallicity relation obtained by Edvardsson et al. (1993)
shows a considerable
scatter at almost all ages and hence any chemical evolution models can easily
reproduce this relationship which does not constitutes 
a tight constraint anymore (figure 1a and 1b).

 
The observed scatter in these data could be real
as discussed by Fran\c cois and Matteucci (1993)
and Matteucci (1994), and it could be the consequence
of the overlapping of different stellar populations observed in the solar
vicinity, possibly caused by orbital diffusion.
Other explanations are given in Edvardsson et al. (1993) (see also
Pagel and Tautvaisiene 1995 and references therein).

\subsection{Stellar metallicity distribution}
An important constraint on chemical evolution models is
the metallicity distribution  of  G-dwarfs for the solar vicinity.
The G-dwarf metallicity distribution is representative of the chemical
enrichment of the Galaxy, since these stars have lifetimes greater
than or equal to the age of the Galaxy and hence can provide 
a complete record of the chemical evolutionary history.
Until now, all the authors made use of the G-dwarf metallicity
distribution published by Pagel and
Patchett (1975) and revised by Pagel (1989) and Sommer-Larsen (1991).
Recently, two different groups using new observations and up to date
techniques, published new data on the G-dwarf distribution (Rocha-Pinto and
Maciel 1996, Wyse and Gilmore 1995). 
The basic differences are in the new 
adopted catalog, namely, the Third Gliese Catalog, and in the calibration 
used to determine the metallicity. This calibration is based on Str\"omgren 
photometry which allows a more reliable estimation of the metallicity 
than does the one based on UBV photometric system as in Pagel and Patchett 
(1975).
As can be seen in figure 2, 
these new data are very similar to each other but very
different from the previous ones.
In particular, the new data show a well-defined peak in metallicity
(between [Fe/H]=-0.3 and 0.), which was not evident in the previous data.


\subsection{Relative abundances}
Another important constraint is the behaviour of the $\alpha$
-to-iron
ratio as a function of iron.
While the $\alpha$-elements (O, Ne, Mg, Si, Ca etc..) are  produced only
by type II SNe (which have high mass progenitors with short life time),
most of the iron is produced by type Ia SNe, which are believed to be the 
result of the explosion of C-O white dwarfs in binary systems. Iron
release from type Ia SNe begins not before several $10^7$ years after
the birth of a star generation, and the bulk of restitution takes up to
some Gyrs, depending on the assumptions on the binary system characteristics,
explosion mechanism and 
star formation rate (SFR).
Therefore, the delayed arrival of the iron
produced by type Ia SNe is responsible for the observed decrease in the
[$\alpha$/Fe] ratio as a function of the iron abundance in the solar vicinity
(Greggio and Renzini 1983, Matteucci and Greggio 1986, MF89).
Recently Gratton et al. (1996) obtained an uniform data set for iron and
oxygen abundances in field stars based on the original equivalent widths
reported in a serie of papers (Gratton and Sneden 1991, Gratton and Sneden
1994, Tomkin et al. 1992, Sneden et al. 1991, Kraft et al. 1992,
Nissen and Edvardsson 1992, Zhao and Magain 1990). In total, the authors
determined [O/Fe] ratios for about 160 stars. These data are presented
in figure 3. Such abundances constitute an important constraint to 
the chemical evolution models. 
As shown in Gratton et al. (1996), a plot of [Fe/O] vs. [O/H] shows that
[Fe/O] increases by $\simeq$ 0.2 dex while the [O/H] ratio holds constant
during the transition from the thick to the thin disk phase. This clearly
indicates the existence of a gap in the star formation rate during the 
transition between these two phases.
In this work we also compare the model
predictions for the abundance ratios of other elements 
with respect to iron with the available observations (section 4).


\subsection{Abundance gradients}
The existence of abundance gradients in the Galaxy 
(and in the disk of external galaxies) has been
one of the fundamental constraints for chemical evolution models. Observations
of H II regions (Shaver et al. 1983) and planetary nebulae of type II,
which can be considered tracers of the abundance gradients
in the ISM (Maciel and K\" oppen 1994, Maciel and Chiappini 1994), 
show a negative abundance gradient for oxygen of the order of -0.07 dex/kpc. 
Planetary nebulae also show gradients for other
elements such as Ne, Ar and S, in the galactocentric
range 5-12 kpc, similar to that
of oxygen.
All previous chemical evolution models tried to reproduce this feature.
However, at present, the existence of
such abundance gradients is controversial. Recently Kaufer et al. (1994)
obtained spectra and abundances of B-type
main sequence stars in young open clusters and in H II regions, 
in the galactocentric distance range 7-16 kpc
and they found a flat 
gradient for O and N in this region. The authors stressed that early type B 
stars on or near the main sequence
should be good indicators for the present 
composition of the interstellar medium
since they have the advantage of being still located near their birthplace,
they did not underwent through mixing processes and
are bright enough to be observed at great distances. 
Based on these data, they concluded that
there is no gradient at least for $R > 6$ kpc, where $R$ is the galactocentric
distance.
As suggested by the authors, this apparent controversial result of gaseous 
nebulae and B stars on radial 
abundance gradients could be solved if the data of Shaver et al. (and
also those on PNe) are interpreted as showing a gradient only
in the inner Galaxy ($R<R_{\odot}$) and a flat profile at larger distances.
This was also suggested by Vilchez and Esteban (1996), 
who studied the chemical 
composition of H II regions at larger
radii and, despite the scatter, he also found a flat oxygen gradient
for galactocentric distances greater than that of the sun. 
Recently, a result reported by Simpson et al. (1995), 
based on infrared abundance
measurements in H II regions, also shows that the radial abundance gradients
seem to be different in the outer and inner parts of the Galaxy.
On the other side, recent work by Liu et al. (1995) shows
that the {\it standard} method of abundance determination in gaseous
nebulae, based on forbidden-line analysis, may underestimate the abundances
by a factor of two to three. How this can affect the abundance gradients
obtained from gaseous nebulae is still not clear.
Given these controversial results, the observed
abundance gradients should be considered with caution.
\par

The observational values for the other constraints in the solar neighbourhood,
namely the relative number of thin-disk and metal poor stars, 
the supernova rates
at the present time, the present day gas fraction, the solar abundances,
the enrichment of helium relative to metals during the Galactic lifetime
($\Delta Y/\Delta Z$), the present time star formation and infall rates
are shown in tables 2 and 3 together with the model results (section 4).

\section{The Model}

\subsection{Model Assumptions}

Our model is based on the MF89
model, but the (halo)-thick-disk and thin disk evolutions
occur at different rates due mostly to different accretion rates.
As in the MF89 model the Galactic disk is
approximated by several independent rings, 2 kpc wide, without
exchange of matter between them. The basic equations are the 
same as in MF89:

\vfill\eject 

\begin{eqnarray}
{d \; G_i(r,t) \over dt}  = 
  - \; X_i(r,t) \; \Psi(r,t) \; +  
\int_{M_L}^{{M_B}_m}\Psi(r, t-\tau_M) \; (Q_M)_i \; \Phi(M) \; dM 
 + \nonumber \\ A
 \int_{{M_B}_m}^{{M_B}_M} \Phi(M_B) 
\biggl[\int_{\mu_m}^{0.5} f(\mu) \;
\Psi(r, t-{\tau_M}_2) \; ({Q_M}_1)_i(t-{\tau_M}_2) \; d\mu \biggl] \; 
dM_B \nonumber \\
 +  
(1-A) \int_{{M_B}_m}^{{M_B}_M} \Psi(r,t-{\tau_M}_B) \; 
({Q_M}_B)_i(t-{\tau_M}_B) \; \Phi(M_B) \; 
dM_B 
\; \nonumber \\ +
 \int_{{M_B}_M}^{M_U} \Psi(r,t-\tau_M) \; (Q_M)_i(t-\tau_M) \; \Phi(M) \; dM
+ 
(X_i)_{inf} \; {d \; G(r,t)_{inf} \over dt} 
\end{eqnarray}

\noindent
where
$G_i(r,t)=[\sigma_g(r,t) \; X_i(r,t)] / \sigma(r,t_G)$ with $G_i(r,t)$ being
the normalized surface gas density in the form of the element $i$,
$\sigma(r,t_G)$
being the present time total surface mass density, $\sigma_{g}(r,t)$
the surface gas density and $X_i(r,t)$ the abundance by mass of an element $i$.
For a detailed description of the meaning of each symbol
in equation (1) see Matteucci and
Greggio (1986).
The two main differences between the present
model and the MF89 model
are the rate of mass accretion and the rate of star formation.
This model assumes two distinct infall episodes: 
the first during which the thick disk is formed 
and the second, delayed relatively to the first, during which the 
thin disk forms. 
The
reader should keep in mind that in this work we do not try to model
explicitly the galactic halo
and that the halo evolution 
(i.e. very low metallicity stars) is included
in the evolution of the thick disk (see section 1).
The thin disk starts forming roughly at
the end of the thick disk phase. In this model,
the material accreted by the Galactic thin disk comes
mainly from extragalactic
sources and this is the fundamental difference between our present
model and the one by MF89. The extragalactic sources could be,
for instance, the Magellanic Stream or a major accretion episode
(see Beers and Sommer-Larsen and references therein).
A long timescale for the formation of the disk
was also suggested by chemodynamical models (Burkert et al. 1992)
which predict a delay of several billion years between the formation
of the last stars in the halo and the appearance of a disk.

The new expression for the rate of mass accretion in each shell is given by:

\begin{equation}
{d \; G_i(r,t)_{\inf} \over dt} = 
 {A(r) \over \sigma (r,t_G)} \; 
(X_i)_{\inf} \;  
e^{-t/\tau_T} + {B(r) \over \sigma (r,t_G)} \; (X_i)_{\inf} \;  
e^{-(t-t_{\max})/\tau_D} 
\end{equation}

\noindent
where
$G_i(r,t)_{\inf}$ is the normalized surface gas density 
of the infalling material in the form of the element $i$,
${(X_i)}_{\inf}$ gives the composition of the infalling gas, which we 
assume to be primordial,
$t_{\max}$ is the time of maximum gas accretion onto the disk,
and $\tau_{T}$ and $\tau_{D}$ are the timescales for the mass accretion in 
the thick disk and thin disk components, respectively.
These are the two really free parameters
of our model and are constrained mainly by comparison with the observed
metallicity distribution in the solar vicinity. The $t_{\max}$ value
is chosen to be 2 Gyrs and roughly corresponds to the end of the thick
disk phase.
The quantities $A(r)$ and $B(r)$
are derived by the condition of reproducing the current total 
surface mass density distribution in the solar neighbourhood.
The current total surface mass distribution is taken from Rana (1991). 
For the thin disk we assume
a radially varying $\tau_{D}(r)$ which implies that the inner parts of the
thin disk are built much more rapidly than the outer ones. In other words, 
we are dealing with an 
inside-out picture, as suggested by
previous models (Larson 1976, MF89,
Burkert et al. 1992). 
According to
Burkert et al. (1992) the reason for an inside out picture of the disk
formation is the strong dependence of the disk evolution on its surface
density which is higher at inner regions, whereas the reason proposed by
Larson (1976) was an infall timescale which increases with the galactocentric
distance.

The adopted radial dependence of $\tau_D$ is:

\begin{equation}
\tau_{D}(r)=0.875r-0.75
\end{equation}

The expression (3) was built in order to obtain
a timescale for the bulge formation ($R < 2 $ kpc) of 1 Gyr,
in agreement with the results of Matteucci and Brocato (1990),
and a timescale of 8 Gyr at the solar neighbourhood,
which best reproduces the G-dwarf metallicity distribution
(we are adopting $R_{\odot} = 10$ kpc).
We assume also that the e-folding time of the
thick disk infall rate is $\tau_T=1$ Gyr, 
which roughly coincides with the appearance of the
first type Ia SNe and which leads to a good agreement with the 
available constraints (section 4).

For the star formation rate we adopt the same expression as in MF89,
which has the same functional form for both thick and thin disk phases:

\begin{eqnarray}
\Psi(r,t)=\tilde{\nu} \; \biggl[{\sigma(r,t) \over \tilde\sigma(\tilde r,t)}
\biggl]^{2(k-1)} \; \biggl[{\sigma(r,t_G) \over \sigma(r,t)}\biggl]^{(k-1)} \;
G^{k}(r,t) \nonumber \\
\Psi(r,t)=0. \,\,\,\, when \,\,\,\, \sigma_g(r,t) \leq 7M_{\odot}pc^{-2}.
\end{eqnarray}

\noindent
where $\tilde\nu$ is the efficiency of the star formation rate expressed
in units of $Gyr^{-1}$, $\tilde\sigma(\tilde r,t)$ is the total surface
mass density at a particular distance $\tilde r$ from the Galactic center
which is taken, as in Chiosi (1980) and MF89, equal to 10 kpc.
As it will be discussed in section 4, the $k$ and $\tilde\nu$ 
parameters of the
star formation rate are not arbitrarily chosen. For the models to be
in agreement with the observational constraints these two parameters must
be restricted to a very small range of values (see section 5).

In the majority of chemical evolution models the star formation rate
is assumed to depend on the surface gas density: 
$\Psi = A \; {\sigma_g}^k$, where $k$ varies from 1 to 2 (Kennicutt 1989).
According to Dopita and Ryder (1994), both a law of the form
of $\Psi = A \; {\sigma_g}$, and a star formation dependent on the
angular velocity at the radial point considered, e. g., 
triggered by the spiral arms, seem to be excluded by observations.
Instead, they suggest that a star formation which depends either on 
the surface gas density or on the total surface mass density, similar to 
the one presented in equation (4), is in better agreement with
the observations.
Subsequent models have adopted different formulations for the
coefficient A and its radial variation.
One of them, as proposed originally by Talbot and Arnett (1975)
and adopted by other authors (cf. Chiosi 1980, Dopita 1985, MF89)
was a dependence on the total surface mass density. 
This can be understood as a way to take into account the {\it local
environment} or, in other words, to consider the fact 
that the more important is 
the gravitational potential, the easier is for the gas to collapse into stars. 
The consequences of the adoption of a star formation rate given by equation
(4) will be discussed in section 4.
This model also considers a star formation
threshold of $7 M_{\odot} pc^{-2}$ (Gratton et al. 1996 and references
therein), which means that below a critical surface gas density there is no
star formation. This threshold is suggested by observations relative to 
the massive star formation in external galaxies (Kennicutt 1989).
The physical reason for a threshold in the star formation is related
to the gravitational stability according to which, below a critical
density the gas is stable against density condensations, and consequently
the star formation is suppressed, the actual critical value depending
also on the rotational properties of the galaxy. The resulting star formation
rate is shown in figure 4, for model A, where the threshold effect is clear
during the thick disk phase and also at the end of the thin disk evolution.


As in MF89 the adopted initial mass function (Scalo 1986) is assumed
to be constant in space and time.

The input parameters ($\nu_T$, $\nu_D$, $k_T$, $k_D$,
$\tau_T$, $\tau_D$) are summarized in Table 1 where model A corresponds to
our best model. 
The subscripts T and D refer to the thick and thin disk, respectively.
All the models listed in Table 1 have the same threshold
in the star formation rate.
The model predicts the abundances of the elements
H, D, $^{3}He$, $^{4}He$, C, O, N, Ne, Mg, S, Si, Ca, Fe, Zn, Cu, 
n-rich 
elements, the gas density, total mass density, star density, rate of 
type I and type II SNe as functions of time and galactocentric distance.

\subsection{Nucleosynthesis Prescriptions}

One of the most important ingredients for chemical evolution models is
the nucleosynthesis prescription and the computation of 
stellar yields. As in MF89, we
adopted the Talbot and Arnett (1975) formalism.

For the D and $^{3}He$ nucleosynthesis we adopted very recent results
taken from Dearborn et al. (1995). The authors have computed new stellar
models which give the production and destruction of $^{3}He$ and $D$,
for stellar masses in the range of 0.65 to 100 $M_{\odot}$. They tried
several non-standard models in order to overcome the problem of the
excessively large $^{3}He$ production in low and intermediate mass
stars which leads to an overabundance of such element in the interstellar
medium, as predicted by almost all the available chemical evolution models
(Tosi 1996).
In the present work we adopted one of such prescriptions, 
characterized by the 
following values of  $6.2 \; 10^{-5}$ and $3.5 \; 10^{-5}$
for the primordial abundances by mass of $D$ and $^{3}He$,
respectively, and $2.1 \; 10^{-5}$ for the main sequence
$^{3}He$ abundance.

For Zn and Cu we adopted the prescriptions given by the best
model of Matteucci et al. (1993).

\subsubsection{ 
Low and Intermediate Mass Stars: ($0.8 \leq M/M_{\odot} \leq 8$)}

{\bf Single Stars} : 
The single stars in this mass range contribute to the Galactic enrichment
through planetary nebula ejection and quiescent mass loss. They enrich the
interstellar medium mainly in He, C and N. For these stars, which end
their lives as white dwarfs, we adopt the prescriptions
of Renzini and Voli (1981), for the case $\alpha =1.5$ and $\eta =0.33$.
More recent calculations for the low and intermediate mass stars contribution
have been computed by Marigo et al (1996). These calculations were
made for a scenario which considers the overshooting process in which the 
$M_{up}$ corresponds to 4 $M_{\odot}$. In this work we opted for 
the stellar evolution scenario without overshooting.

\smallskip
\noindent
{\bf Type Ia SNe}:
Type Ia SNe are thought to originate
from C-deflagration in C-O white dwarfs in binary systems.
The type Ia SNe contribute to a substantial amount of
iron ($\sim 0.6 M_{\odot}$ per event) and to non-negligible quantities
of Si and S. They also contribute to 
other elements such as O, C, Ne, Ca, Mg and Ni, but in negligible amounts
when compared with the masses of such elements ejected by type II SNe.
The adopted
nucleosynthesis prescriptions are from Thielemann et al. (1993)
and they are an updated version of those utilized in MF89 (Nomoto et al. 1984).
The differences between these new yields and 
the old ones are mainly for Ne and Mg, but small enough to have no 
influence on the model predictions.
As in MF89, type Ib SN 
events are assumed to be a half of the total number of type I SN events
and to contribute only to the iron abundance ( $\sim 0.3 M_{\odot}$
per event, MF89). 
\medskip
\subsubsection{Massive Stars: ($8 < M/M_{\odot} \leq 100$)}

These stars are the progenitors of type II SNe. For this range of masses we
adopt up to date stellar evolution calculations by Woosley and Weaver (1995)
for the following elements: $^{4}He$, $^{12}C$, $^{13}C$, $^{14}N$, $^{16}O$,
$^{20}Ne$, $^{24}Mg$, $^{28}Si$, $^{32}S$, $^{40}Ca$ and $^{56}Fe$.
The major advantage of these new calculations is 
that explosive nucleosynthesis is taken into account. This was not
the case in the yields adopted by MF89 (Woosley and Weaver 1986)
relative to pre-supernova models which required a 
very uncertain correction
factor to take into account the fact that part of Si and S is transformed
into iron by explosive nucleosynthesis and that part of some elements
falls back onto the collapsing core before the explosion starts (MF89).
The second advantage is that in the new calculations the yields for 
$^{13}C$ and $^{14}N$ are also included.
There are important differences
between these new prescriptions for the massive
nucleosynthesis yields and those previously reported by Woosley and Weaver
(1986). 
In particular, for the elements He, Ne and Fe the new yields are higher. 
On the other hand, the new values for Mg, Si, S and Ca are lower than the 
previous ones (see Gibson 1996 for a comparison between stellar yields
from different sources). 
We present in figure 5 the masses ejected in form of various elements,
weighted by the IMF as a function of the initial mass of the progenitor 
star. Comparing this figure with figure 4 of MF89 we see that the 
major differences arise in the Si and S abundances for which the 
explosive nucleosynthesis is very important, as discussed before.
The effect of these new yield
prescriptions on models predictions will be discussed in section 4.


\begin{table}
\caption{Input Parameters for the solar neighbourhood}
\smallskip
\label{}
\begin{tabular}{ccccccr}
\hline
 & & & & & & \\
Model & ${\tilde\nu}_T$ & $k_T$ & ${\tilde\nu}_D$ 
& $k_D$ & $\tau_T$ & $\tau_D$ \\
 & ($Gyr^{-1}$) & & ($Gyr^{-1}$) & & ($Gyr$) & ($Gyr$) \\
\smallskip
 & & & & & & \\
\hline
& & & & & & \\
A & 2.0 & 1.5 & 1.0 & 1.5 & 1.0 & 8.0 \\
B & 2.0 & 1.5 & 1.0 & 1.5 & 1.0 & 5.0 \\
C & 2.0 & 1.5 & 1.0 & 1.5 & 0.1 & 5.0 \\
D & 2.7 & 2.0 & 1.0 & 1.5 & 1.0 & 5.0 \\ 
& & & & & & \\
\hline
\end{tabular}
\end{table}

\section{Results}

We ran a large number of models for the solar vicinity and the whole disk, 
varying the 
star formation rate parameters (i.e. $k$ and $\nu$) and also the 
timescales for the thick and 
thin disk formation (i.e. $\tau_{T}$ and $\tau_{D}$). 
A positively surprising result is
that only very few combinations of such parameters lead to an agreement
with the considered set of observational constraints. Table 2 presents
the results of the models of table 1 compared with the current observational
quantities, for the solar vicinity. We remind that model A is our best model.

\subsection{The relative number of metal poor and metal rich stars}

As discussed by Matteucci et al. (1990) and references therein, 
attempts to form a disk out of 
already enriched halo gas and with an initial mass function constant in space
and time overestimate the fraction of halo stars, although
these models reproduce the distribution of the G-dwarfs (MF89). 
It was then suggested that
infall of primordial extragalactic gas coupled with infall of
halo gas could overcome this problem.
It is clear from our models that in order to 
obtain a reasonable number of metal poor stars
and simultaneously fit the metallicity distribution of the solar vicinity
is necessary to decouple the evolution of the (halo)-thick-disk 
from that of the thin disk. As already pointed out in section 1, this decouplement
is also required from the observational point of view (Wyse and Gilmore 1992,
Beers and Sommer-Larsen 1995).

The observed value of the relative number of disk and halo stars
suggested by Pagel and Patchett (1975) is 3 \%. However, this value
is underestimated as it is based on stars observed only up to a certain
scale height above the Galactic plane ($\simeq 1 \; kpc$ for G-dwarf stars).
Matteucci et al. (1990) estimated that the number of stars
contained in a cylinder around the solar neighbourhood with
practically an infinite height above the Galactic plane,
which is the actual quantity predicted by the chemical
evolution models, should be a factor 3-4 bigger than the estimated by
Pagel and Patchett (1975), which means $\simeq$ 10 \%.
In the definition of this quantity we have defined (halo)-thick-disk stars all those with
[Fe/H] $<-1.0$ and thin disk stars all the others.
However, these definitions should be taken with care in view of the fact that
part of the low metallicity stars could have been accreted from satellites
galaxies (Gratton et al. 1996), and therefore would not appear in our
predictions.
Our best model (model A)
predicts a fraction of low metallicity stars in the range $6-10\%$, the 
exact value depending on the assumed duration of the thick disk phase
(1 to 2 Gyrs).
Model C, which assumes a timescale for the thick disk formation of 0.1 Gyrs,
predicts a very high value for this fraction because in this case
the thick disk evolution is faster, and more stars are formed during this
phase. Models B and D predict very similar ratios 
which are smaller than in model A. In fact, 
the timescale for thin disk formation in models B and D is smaller than
in model A and therefore they predict more thin disk stars and consequently
a smaller ratio between metal poor stars and thin disk stars.

\begin{table*}
\caption{Current predicted and observed quantities for the solar
neighbourhood}
\smallskip
\label{}
\begin{tabular}{cccccr}
\hline
 & & & & & \\
  & A & B & C & D & Observations \\
 & & & & & \\
\hline
 & & & & & \\
metal-poor/total 
& 6-13 \% & 5-10 \% & 17 \% & 5-10 \% & $\simeq$ 10 \% \\
stars & & & & & \\
SNI $century^{-1}$ & 0.29 & 0.37 & 0.37 & 0.38 & 0.17-0.7 \\
 & & & & & \\
SNII$century^{-1}$ & 0.78 & 0.53 & 0.54 & 0.55 & 0.55-2.2 \\
 & & & & & \\
SNII/SNI & 2.7 & 1.4 & 1.5 & 1.4 & 3.1 \\
 & & & & & \\
$\Psi(R_{\odot},t_{now})$ & 2.64 & 2.38 & 2.34 & 2.38 & 2-10  \\
$M_{\odot} \; pc^{-2} \; Gyr^{-1}$ & & & & & \\
 & & & & & \\
$\sigma_g(R_{\odot},t_{now})$ & 7.0 & 7.0 & 7.0 & 7.0 & 6.6 $\pm$ 2.5 \\
$M_{\odot} \; pc^{-2}$ & & & & & \\
 & & & & & \\
$\sigma_g / \sigma_T (R_{\odot},t_{now})$ & 0.14 & 0.12 & 0.11 & 0.12 & 
0.05-0.20 \\
 & & & & & \\
$\dot{\sigma_{inf}}(R_{\odot},t_{now})$ & 1.05 & 0.73 & 0.73 & 0.73 & 1.0 \\
$M_{\odot} \; pc^{-2} \; Gyr^{-1}$ & & & & & \\
 & & & & & \\
$\Delta Y/\Delta Z$ & 1.63 & 1.60 & 1.57 & 1.60 & 3.5 $ \pm $ 0.7 \\
 & & & & & \\
$\Psi(R_{\odot},t_{now})/\langle \Psi \rangle$ & $\sim$ 0.7 & $\sim 0.7$ &
$\sim$ 0.7 & $\sim$ 0.7 & 0.18-3.0 \\
 & & & & & \\
\hline
\end{tabular}
\end{table*}

\subsection{Supernova rates}

The present SN rates in the
Galactic disk were estimated by Van den Bergh (1988) as
$0.6 \; h^2 \; century^{-1}$ for type Ia SNe, $0.8 \; h^2 \; 
century^{-1}$ for type Ib SNe and $2.2 \; h^2 \; century^{-1}$ 
for type II SNe, assuming a total Galactic blue luminosity of 
$\simeq 2.0 \; 10^{10} \; L_{B_{\odot}}$. In table 2 we are 
showing the values for $H=50$ and $100 km\; s^{-1} \;
Mpc^{-1}$ respectively, where an average value was taken for the type
I SNe (Ia + Ib).
Table 2 also shows the ratio between type II and type I SN rates.
We can observe that
our best model is in very good agreement with these constraints.
For the other cases the predicted ratio is
lower than the observed one. This is due, 
basically, to the lower rate of type II SNe predicted in
models B, C and D. 
In fact, model A has a longer timescale for the formation of the thin disk
relative to models B, C and D and,
at present, is accreting more gas thus implying
a higher star formation rate and type II SN rate (Table 2).
This can be seen from the comparison of figures 4 and 6.
This later shows the time evolution of type Ia and II supernovae. The
temporal behaviour of type I SNe is almost independent of the details of the
star formation rate as these objects have longer evolutionary timescales
(see section 4.7).


\subsection{The star formation history}

As already mentioned in section 5, we adopt a threshold in the star
formation rate which is responsible for the behaviour showed in figure
4. According to this threshold, the star formation
stops when a surface gas density of $7 M_{\odot} pc^{-2}$ is reached.
This means that star formation can have an intermittent behaviour regulated
by the surface gas density. In fact, after the threshold is reached and
star formation stops the dying stars continue to restore gas into the
interstellar medium and therefore, soon or later, the surface gas density
will be again above the threshold and star formation will restart.

\subsection{Present-Day Gas fraction}

Our predicted value for the present-day gas fraction (Tosi 1996) 
is in agreement with the observational one for all the models of table 1.
For the best model, however, the predicted present-day gas fraction is
a little higher because of the slower evolution of the thin disk as implied by
a timescale of mass accretion of 8 Gyrs. 

\subsection{Solar abundances}

The solar abundances (by mass) predicted by the best model are compared
with the observed ones (Anders and Grevesse 1989 - AG89) in table 3. These
solar abundances are very similar in all the models
shown in table 1. As already mentioned in section 2, these abundances
should represent the composition of the interstellar medium at the time 
of the formation of the sun, e.g., 4.5 Gyrs ago. 
Since we assume a Galactic lifetime of 15 Gyrs this time corresponds to 
10.5 Gyrs after the Big Bang.
Given the uncertainties involved in observed determinations
we can consider that the model is in agreement with the observed
value inside a factor 2 difference. From table 3 we can see
that only for two elements, namely $^{3}He$ and Mg the model
predictions fail. In the case of $^{3}He$ this was expected as it 
constitutes a problem for almost all the chemical evolution models 
(Tosi 1996, Dearborn et al.
1995). For the Mg this can be attributed to the lower
yield predicted for this elements by Woosley and Weaver (1995).

\begin{table}
\caption{Solar Abundances by Mass}
\smallskip
\label{}
\begin{tabular}{ccr}
\hline
 & & \\
Element & Best Model (A) & Observations (AG89) \\
 & & \\
\hline
 & & \\
$H$ & .731 & .702 \\
$D$ & 4.630 (-5) & 4.80 (-5) \\
$^{3}He$ & 10.01 (-5) & 2.93 (-5) \\
$^{4}He$ & 2.548 (-1) & 2.75 (-1) \\
$^{12}C$ & 1.827 (-3) & 3.03 (-3) \\
$^{16}O$ & 7.278 (-3) & 9.59 (-3) \\
$^{14}N$ & 1.386 (-3) & 1.11 (-3) \\
$^{13}C$ & 4.758 (-5) & 3.65 (-5) \\
$Ne$ & 0.942 (-3) & 1.62 (-3) \\
$Mg$ & 2.48 (-4) & 5.15 (-4) \\
$Si$ & 7.03 (-4) & 7.11 (-4) \\
$S$ & 3.071 (-4) & 4.18 (-4) \\
$Ca$ & 3.95 (-5) & 6.20 (-5) \\
$Fe$ & 1.37 (-3) & 1.27 (-3) \\
$Cu$ & 8.18 (-7) & 8.40 (-7) \\
$Zn$ & 2.44 (-6) & 2.09 (-6) \\
$Z$  & 1.433 (-2) & 1.886 (-2) \\
 & & \\
\hline
\end{tabular}
\end{table}

\bigskip
\subsection{Age-Metallicity Relation}

In figures 1a and 1b we present the age-metallicity relation for our
best model compared, respectively, with two sets of observational data. From 
figure 1a, which presents the data of 
Twarog (1980), Meusinger et al. (1991) and Carlberg et al. (1985),
one can conclude that the model predictions agree quite
well with the observational data. However, from the second set of data
(Edvardsson et al. 1993), although there is an overall trend of decreasing
mean metallicity with increasing age, a large scatter is present and 
we consider of fitting only the mean relation. 
According to Nissen (1995), however, the data of Edvardsson et al. should
not be used to determine the mean age-metallicity relation in the solar
neighbourhood as the sample of stars is biased with respect to [Fe/H],
in the sense that some metal rich stars have been excluded by a temperature
cutoff. The same cutoff seems also to be present in Twarog's data. Given this
situation we cannot consider this constraint as a very tight one.
\medskip

\subsection{Stellar Metallicity Distributions}

A very good agreement is obtained between the  predictions of model A
and the new data for the metallicity distribution of disk stars in the
solar vicinity (figure 7). 
In this figure the observed metallicity distribution includes the thin 
disk stars as well as part of the thick disk ones, since the halo stars 
and the metal weak tail of the thick disk were excluded from the data by 
using a chemical criterion (all stars with [Fe/H] $< -1.2$). 
The same criterion has been used in the model predictions.
The data have been corrected by the factor $f$, as defined by Sommer-Larsen
(1991) in order to take into account the vertical scale height effects.

We stress that the
comparison between model predictions and
metallicity distribution is meaningful only when one
considers explicitly the evolution of iron. 
Indeed, this can be done only by those models 
which take into account
the type Ia SNe contribution.
The majority of chemical evolution models consider IRA and 
does not take into account the type Ia SNe contribution to 
the chemical enrichment of the Galaxy.
Thus, when comparing theoretical predictions with 
observed chemical properties of our Galaxy, they are forced to
use oxygen rather
than iron. 
The solution
often adopted in previous papers was to convert the observed
[Fe/H] into [O/H] using a fit to the [O/Fe] versus [Fe/H] 
relation for solar vicinity stars, in order to do
this conversion.
Such a procedure is quite uncertain and if adopted in the framework of the
present model produces quite different results from those obtained
just comparing the real data ([Fe/H]) with the predicted [Fe/H].

The metallicity distribution
of metal poor stars constitutes another constraint. This metallicity
distribution varies from author to author according the adopted
selection criteria. According to Beers and Sommer-Larsen (1995)
 it is unlikely
that these distributions are not a mixture of halo and thick disk
populations whatever the criteria used, since the properties of both
components overlap one another. 

However, we are interested not to produce a fit of the metallicity distribution
of metal poor stars but only to confirm that our models can predict
a reasonable number of stars within a range in metallicity similar to the
observed one. This was the case of model A, which predicts 
a reasonable number of stars with metallicities in the range
-3 $\leq$ [Fe/H] $\leq$ -1. 
As an example model C which
has $\tau_T$ equals to 0.1 Gyr predicts very few stars with metallicities
[Fe/H] $<$ -2 in contradiction with observations (Ryan and Norris 1991,
Schuster et al. 1993, Laird 1995, Beers and Sommer-Larsen 1995).



Our best model (model A), 
which fits the disk metallicity distribution and predicts a number of
stars which is in agreement with the metalicity distribution of metal
poor stars,
has a star formation in the thick disk more efficient than that in the 
thin disk
($\nu_T=2.0$ and $\nu_D=1.0 Gyr^{-1}$). However the exponent
for the star formation rate is the same for the two phases
($k_T=k_D=1.5$). We also ran models which have a higher star
formation exponent in the thick disk compared
to that in the thin disk 
and the same efficiency for the SFR, but they
did not produce a good fit to the oxygen/iron relation.
It should be noted, however, that these parameter values
must also be in agreement with all the other constraints for the chemical
evolution models. For instance, a very high SFR in the thick disk would
produce a lot of metal-poor stars leading to a too 
high ratio of low-metallicity/thin-disk stars.

Model A predicts a 
timescale for the thick and thin disk formation
of, respectively, $\leq$ 1 Gyr and 8 Gyr.
Timescales of this order have been suggested also by the results of 
chemo-dynamical models of  Burkert et al. (1992) and recently by
Yoshii et al. (1996).
The $\tau_{T}$ value cannot be much less than 1 Gyr because in such a
case the model would predict a metallicity distribution for 
metal-poor stars which is shifted towards a metallicity higher than
the observed one, as a consequence of the faster evolution
of the thick disk (model C).
On the other hand, a value greater than 1 Gyr,
would produce a model which does not agree with the observed behaviour
of the oxygen to iron ratio (i.e. the long plateau in [O/Fe])
and which also predicts too many metal poor
stars in the disk (G-dwarf problem). 
Concerning the timescale for the (halo)-thick-disk formation,
the high value for $\tau_{D}$
is necessary to fit the new observed G-dwarf distribution for disk
stars which shows a much more pronounced peak for intermediate metallicities
than the previous one.
Models with $\tau_{D} \leq$ 6 Gyr lead to 
a poor agreement with the metallicity distribution by Rocha-Pinto and
Maciel (1995).
This explains the larger timescale obtained by the present model compared
with that found by MF89, which was 3-4 Gyrs, and also the higher
exponent $k$ of the star formation rate.

\subsection{The [O/Fe] versus [Fe/H] relation}

The distribution of the abundance ratios 
represents an important constraint for chemical evolution models.  
In fact, they are less dependent on model parameters
than absolute abundances, since depend
essentially on the nucleosynthesis yields.
Figure 3 shows the oxygen/iron ratio with respect to iron relative
to the sun. Our best model (Model A) is represented by the solid line in this
figure.
Although our models 
predict an overabundance of oxygen for
the halo stars, the predicted [O/Fe] is lower than the observed mean value.
This
is a consequence of the high iron yield for type II SNe
taken from Woosley and Weaver (1995).
This fact indicates that the iron yields
from type II SNe should be lower, as also discussed
by Timmes et al. (1995).

Figure 8 shows the star formation rate for a model (model C)
with a timescale
for the thick disk shorter than in model A. 
It can be seen that in models with a short thick disk timescale
the gap in the star formation which arises between the thick and thin
disk phases
lasts longer (compare figure 4 and figure 8). This is a consequence
of the faster consumption of gas together with the
threshold in the star formation
rate, and in a such case the model predicts a looping in the 
oxygen to iron ratio as function of metallicity (dotted line in figure 2).
The looping predicted by the two infall model,
also suggested by observations (see Gratton et al. 1996 and references
therein),
can be understood as to be caused by the strong dilution of the interstellar
medium abundances at the moment of maximum infall onto the disk coupled to
a sudden decrease of the SFR at the end of the halo-thick disk phase.
The dilution lowers the absolute abundances but 
does not affect their ratios, whereas a gap in the SFR would lower the oxygen
abundance (whose progenitors are mainly the type II SNe), but not that
of iron which continues to grow due to the contribution of the type Ia SNe.
However, the star formation gap lasts for a very short time 
because of the gas which is setting onto the disk at a high rate, and so the
decrease in 
the [O/Fe] ratio is much smaller than the decrease in the iron abundance,
as can be observed in figure 3.
However, since we need relatively long timescale (1 Gyr) for the halo
and thick disk formation our best model does not present a very
pronounced star formation gap
(see fig. 4). The net effect on the [O/Fe] versus [Fe/H] relationship is
that the oxygen overabundance (as well for the other elements) does not
remain constant until the disk formation ([Fe/H] $\simeq$ -1), but starts
to decrease slowly from lower metallicities. This can be understood as 
a non negligible contribution of type Ia SNe already in the thick disk phase.


We have also computed models with delayed infall but without
the star formation threshold and with
an artificially imposed gap on the SFR at the end of the thick disk phase.
In this case we have seen that the star
formation in the thick disk should be almost of the same order
as that in the thin disk,
in order to avoid an overproduction of 
both oxygen and iron by the type II SNe in the thick
disk phase, 
resulting in a solar ratio in poor agreement with observations,
which, in turn, gives extremely high [O/Fe] ratios for $[Fe/H]<-1$. 
A better agreement 
between observational constraints and theoretical predictions is achieved
when the SFR in the (halo)-thick-disk is 
more efficient than that of thin disk, 
as also suggested by Larson (1976).

\subsection{The [X/Fe] versus [Fe/H] relation for the other elements studied}

Figures 9 a,b,c,d show the predicted abundances of $\alpha$-elements
(Mg, Si, S and Ca) 
relative to iron as functions of [Fe/H]
compared with observational data. 
Generally, although the spread in the data 
is due mainly to the fact that the data
originate from different sources, the agreement is acceptable. There
is, as already mentioned, a general problem with the
iron yields from supernovae of type II which is probably too high.
However, the agreement between predictions and observations is generally quite satisfactory, perhaps with the exception of S.


In figure 10 (a, b) the model predictions for Zn and Cu are
compared with the observations. For Zn and Cu there is a marginal agreement,
with the curves which lie below the observational data. This is
also a consequence of the high iron yield predicted by the Woosley
and Weaver (1995) stellar evolution models.
An extended discussion about these elements is present by Matteucci et al.
(1993).


Figure 11 shows the [C/Fe] versus [Fe/H] relation. For this element, which
is mostly produced by low- and intermediate-mass stars,
as iron, we should expect a constant and solar ratio. 
As in the MF89 model, 
the predicted [C/Fe] ratio shows 
a small bump around $[Fe/H] = -1.0$, which
should be due to the assumption of not fully consistent nucleosynthesis
yields in the ranges of massive and low- and intermediate- mass stars.
On the other hand, the observational data for this element show a considerable scatter.
Our predictions for C seem also to predict a positive trend for metallicities
smaller than about [Fe/H] $\simeq$ -2.


Figure 12 shows  the behaviour of nitrogen. 
For this element the data also seem to indicate a big scatter.
Moreover, as discussed in Timmes et al. (1995), 
in the data of Laird (1985) there is a zero-point 
correction factor of $\sim +0.65$ dex. All of this makes
the situation for nitrogen quite uncertain and more 
and better data are required.
Our predictions show  an initial steep increase of the [N/Fe] ratio due to the fact that the first N which is produced originates from massive stars as a
secondary element. Moreover, the nitrogen produced in intermediate
and low mass stars is also mostly secondary, 
although some primary N can be produced in massive intermediate mass stars
(Renzini and Voli, 1981), but the exact amount is quite uncertain.
A higher [N/Fe] ratio at low [Fe/H] could be obtained only by assuming
that N in massive stars has a primary origin 
(Matteucci 1986; Timmes et al. 1995). 
Primary N in massive stars 
seems to be required also to explain the abundance pattern in 
damped Ly-$\alpha$ systems
(Matteucci et al. 1996).


Figure 13 shows the predictions for the log(X/O) versus O/H relationship
for S, N, C and Ne. The Ne/O and S/O ratios as function of
oxygen are almost constant, an expected fact since  these elements are produced
in the same stars as oxygen. The growth
of the N/O ratio is showing again the 
assumed secondary origin of nitrogen in massive stars.
The increase of the C/O ratio with oxygen
is due to the fact that carbon is mostly
produced in low- and intermediate-mass stars whereas most of the oxygen
comes from the massive star range.


\subsection{Radial Profiles}

Here we discuss the predicted radial properties of our best model A.

Figure 14 shows the predicted [O/Fe] versus [Fe/H] relations for three different galactocentric distances (4, 10 and 14 kpc). As first pointed out by
Matteucci (1992), the slope of the [O/Fe] in the thin disk metallicity range steepens with increasing galactocentric distance as due to the differential evolution of the thin disk. 
Here, another effect of this differential evolution 
appears, namely 
a loop in the [O/Fe] develops at 4 kpc. This is due to the higher maximum
infall at this galactocentric distance relative to larger distances.

As shown by Rana (1991), the distribution of
neutral and molecular hydrogen in the Galaxy is poorly known and 
independent observations do not agree with each other. 
In figure 15 we compare 
our model predictions for the radial gas distribution with the observed
one (taken from Rana 1991, table 3). Given the uncertainties in the
observational data, we can say that the model is in agreement
with the observed radial profile. Some disagreement, however,
seems to arise for
galactocentric distances smaller than 4 kpc, where the model predicts
much more gas than is observed. A possible solution to this problem
could be the existence of a bar in our Galaxy
which would strip the gas and push it into the Galactic center.
For the outer radii the model predicts a flat distribution but observational
data are not available.



The observed
radial dependence of the present time star formation
rate is also very uncertain.
Usually pulsars,
supernovae remnants and Ly continuum photons are used as tracers of the present
star formation rate across the disk. However, as discussed by Kennicutt 
(1989) for the purpose of chemical 
evolution model calculations, a global star formation
rate process should be taken into account instead of local ones.
So it is not clear if this comparison is meaningful 
given the
large uncertainties involved. One example of this kind of problem is 
the apparent lack of correlation, found by Kennicutt, between the radial
distribution of the global star formation rate and molecular clouds in external
galaxies.

Table 4 shows the gradients predicted by our best model
for the inner ($4 \leq R(kpc) \leq 10$)
and outer ($10 \leq R(kpc) $) regions of the Galactic disk as 
functions of time,  for O, Fe, N, S and Ne.
The same is shown for D, $^{3}He$ and $^{4}He$ in table 5.
The quoted errors represents the confidence of the theoretical bestfits.
The gradients are shown also as functions of Galactic ages,
namely
15, 12, 10.5 ($t_{\odot}$), 7 and 2 ($t_{infall max}$) Gyrs.
Our best model predicts a steeper gradient in the inner regions, which grows
with time and a flatter one in the outer parts. As discussed in
section I, this result seems to be in agreement with the observational
data. 
We note that one of the consequences in
adopting a differential formation of the thin disk
is that we expect different time histories of gradients in the inner
and outer regions.
In particular, the gradients in the outer regions grow quicker than those
in the inner regions but reach very soon a saturation value. This is due
to the fact that in the outer regions the threshold 
value in the gas density is
reached very early, due to the lower gas density always present there, 
whereas in the inner regions such as 4 kpc the threshold is not yet 
reached at the present time.


\small
\begin{table}
\vspace{-1cm}
\caption{Abundance gradients (dex/kpc)}
\smallskip
\label{}
\begin{tabular}{ccr}
\hline
Time (Gyr) & inner region & outer region \\
\hline
O & & \\
15 & -0.032 $\pm$ 0.005 & -0.018 $\pm$ 0.001 \\
12 & -0.023 $\pm$ 0.006 & -0.017 $\pm$ 0.001 \\
10.5 & -0.018 $\pm$ 0.007 & -0.017 $\pm$ 0.001 \\
7 & -0.003 $\pm$ 0.005 & -0.014 $\pm$ 0.001 \\
2 & 0 & 0 \\
Fe & & \\
15 & -0.040 $\pm$ 0.008 & -0.017 $\pm$ 0.002 \\
12 & -0.027 $\pm$ 0.010 & -0.014 $\pm$ 0.002 \\
10.5 & -0.020 $\pm$ 0.009 & -0.013 $\pm$ 0.001 \\
7 & -0.001 $\pm$ 0.007 & -0.010 $\pm$ 0.001 \\
2 & 0 & 0 \\
N & & \\
15 & -0.037 $\pm$ 0.005 & -0.028 $\pm$ 0.002 \\
12 & -0.027 $\pm$ 0.007 & -0.027 $\pm$ 0.002 \\
10.5 & -0.021 $\pm$ 0.008 & -0.026 $\pm$ 0.003 \\
7 & -0.002 $\pm$ 0.007 & -0.022 $\pm$ 0.002 \\
2 & 0 & 0 \\
S & & \\
15 & -0.036 $\pm$ 0.006 & -0.021 $\pm$ 0.001 \\
12 & -0.025 $\pm$ 0.008 & -0.019 $\pm$ 0.001 \\
10.5 & -0.019 $\pm$ 0.008 & -0.019 $\pm$ 0.001 \\
7 & -0.003 $\pm$ 0.006 & -0.015 $\pm$ 0.001 \\
2 & 0 $\pm$ 0 & 0 $\pm$ 0 \\
Ne & & \\
15 & -0.030 $\pm$ 0.004 & -0.019 $\pm$ 0.00 \\
12 & -0.023 $\pm$ 0.006 & -0.020 $\pm$ 0.001 \\
10.5 & -0.018 $\pm$ 0.006 & -0.021 $\pm$ 0.001 \\
7 & -0.004 $\pm$ 0.005 & -0.018 $\pm$ 0.001 \\
2 & 0 $\pm$ 0 & 0 $\pm$ 0 \\
\hline
\end{tabular}
\end{table}
\normalsize

\begin{table}
\caption{Abundance gradients (dex/kpc)}
\smallskip
\label{}
\begin{tabular}{ccr}
\hline
 & & \\
Time (Gyr) & inner region & outer region \\
 & & \\
\hline
 & & \\
D & & \\
15 & 0.026 $\pm$ 0.008 & 0.003 $\pm$ 0.001 \\
12 & 0.009 $\pm$ 0.004 & 0.002 $\pm$ 0 \\
10.5 & 0.005 $\pm$ 0.003 & 0.001 $\pm$ 0 \\
7 & 0 $\pm$ 0 & 0.001 $\pm$ 0 \\
2 & 0 & 0 \\
 & & \\
$^3{He}$ & & \\
15 & -0.027 $\pm$ 0.018 & -0.006 $\pm$ 0.002 \\
12 & -0.001 $\pm$ 0.015 & 0 $\pm$ 0 \\
10.5 & 0.005 $\pm$ 0.011 & 0.003 $\pm$ 0.001 \\
7 & 0.005 $\pm$ 0.004 & 0.004 $\pm$ 0.001 \\
2 0 & 0 \\
 & & \\
$^4{He}$ & & \\
15 & -0.006 $\pm$ 0.001 & -0.002 $\pm$ 0.000 \\
12 & -0.004 $\pm$ 0.001 & -0.002 $\pm$ 0.000 \\ 
10.5 & -0.003 $\pm$ 0.001 & -0.002 $\pm$ 0.000 \\
7 & 0 & -0.001 \\
2 & 0 & 0 \\  
 & & \\
\hline
\end{tabular}
\end{table}

\section{Discussion and Conclusions}
In this paper we presented a model for the chemical evolution of 
the Galaxy
assuming that the formation of thick disk and thin disk
occurred in two main accretion episodes.
New stellar yields and a threshold in the star formation process were also 
considered. From the comparison 
between observational constraints and our theoretical results, we
can conclude that the two-infall model is in good agreement with
most of the observed features of the Galaxy.
Our main conclusions are:
 
\begin{itemize}

\item
Our best model
predicts the abundances of 16 chemical elements well in agreement 
with the observed solar abundances with the exception of $^{3}He$ and Mg. 
The behaviours of relative ratios
as functions of [Fe/H] also agree with the available observational data
although the comparison between theory and observations suggest
to lower the iron yield from type II SNe and to increase the Mg yield
from type II SNe. 
The same conclusion concerning Fe was reached by Timmes et al. (1995)
who adopted the same yields.

\item
The assumption of a 
threshold for the star formation naturally leads to a star formation gap at the end
of the halo and thick disk phases as suggested by 
Gratton et al. (1996), and
prevents an overproduction of oxygen by halo stars.
This star formation gap reflects in a more or less pronounced decrease in the
$[\alpha/Fe]$ ratios around $[Fe/H]=-0.6$ dex, depending on
the duration of the gap itself.

\item
The [$\alpha$/Fe] versus [Fe/H] relations for stars born at different
radii than the solar circle indicate that 
a loop in the [$\alpha$/Fe] develops at small galactocentric distances.

\item
Our best model predicts a stellar metallicity
distribution for the G-dwarfs in the disk
in very good agreement with the new observations
(Rocha-Pinto and Maciel 1996, Wyse and Gilmore 1995).
This constitutes the most important constraint for chemical evolution
models at present and implies a timescale for disk formation much
longer than that for the formation of the halo (and thick disk).

A reasonable agreement is found also when comparing 
the predicted and observed number of very low metallicity stars
(Ryan and Norris 1991, Schuster et al. 1993, Laird 1995, Beers and Sommer-Larsen 1995).
The model predicts a 
timescale of 1 Gyr for the formation of the halo and thick disk
and 8 Gyr for the formation of the thin disk in the solar vicinity.

The agreement with the metallicity stellar distribution and also with the 
fraction of metal poor stars, in the solar neighbourhood,
constitutes an important and new
result being a consequence of, at least, three main assumptions of our model.
First, the decoupling between the rate of gas loss from the
halo-thick disk and that of gas infalling onto the thin disk,
allowing a much longer timescale for the thin disk evolution
as compared to that of the thick disk phase.
Second, the delay in the beginning of the thin disk evolution, which
characterizes a sequential model,
at variance with the models of Matteucci et al (1990),
Pardi and Ferrini (1994) and Pardi et al. (1995)
which adopt the view of a parallel 
evolution of all galactic components (halo, thick and thin disk) 
at different evolutionary rates. 
These models which consider a superposition
of stellar populations,
evolving simultaneously do not best fit the distinct
enrichment histories of the different components at the same time. 
Finally, the third reason, 
is a threshold in the star formation rate which limits
the star formation in the halo phase allowing for the formation
of a number of stars with small metallicities as required by
the metallicity distribution of metal poor stars. 

\item
Our best model requires a more steep dependence of the surface
gas density ($k=1.5$) compared to what was previously reported by MF89
($k=1.1$). It has also been shown that the star formation exponent should
be the same for thick and thin disk phases in order to reproduce the 
observational constraints. However, during the thick disk phase the star formation
should be more efficient and this is obtained by adopting a bigger value
for the efficiency of star formation, the constant $\tilde {\nu_T}$.
Dopita and Ryder (1994) have investigated various models of star formation
in disk galaxies. These authors
showed that the star formation rate in disk galaxies determines their
stellar surface brightness and the surface brightness in $H\alpha$. From a 
star formation rate of the same type of the one used here, but expressed
in a more general form ($\Psi \; \alpha \; {\sigma_g}^m \; {\sigma_T}^n$),
they claim that a satisfactory agreement with the observed properties
of such galaxies is achieved for $1.5<m+n<2.5$. In terms of our formulation,
this means: $1.25 < k < 1.75$, which is in extremely good agreement
with our best model prediction of $k=1.5$. 
Prantzos and Aubert (1995)
investigated the effect of different assumptions about the star formation
rate on chemical evolution models. They tested also the one given by
Dopita and Ryder (1994), for the cases $n=1$, $m=1$ and $n=0.5$, $m=1$ and
concluded, at variance with us,
that a star formation rate given by the expression above, which
depends not only on the gas mass density but also on the total mass density,
did not fit all the required constraints, and adopted instead a star formation
rate given by: $\Psi \; \alpha \; (1/R) \; {\sigma_{g}(R)}^n$ which
reproduces the steep distribution of the star formation rate
given by the observational data.
The authors argued that the discrepancy
between the SFR proposed by Dopita and Ryder (1994) and
the observational constraints was due to a too steep radial dependence 
leading to an overestimation of the amount of gas in the outer parts
of the Galaxy. 
However, given our results, we suggest that Prantzos and Aubert (1995)
did not find an agreement between the star formation proposed by 
Dopita and Ryder (1994) and the observational properties 
because of the mild dependence
on the gas mass density they adopted.

\item
Steeper
abundance gradients for O, N and S are predicted for the inner regions of the Galactic disk ($R < R_{\odot}$) relative to the outer regions. 
This result is in agreement with recent abundance determination in B supergiants (Kaufer et al. 1994)
and with abundance measurements of HII regions (Vilchez and Esteban 1996;
Simpson et al. 1995). 
The gradients obtained by MF89 were generally steeper than the present ones although the difference between the inner and outer regions was already present.
The difference between the present results and those of MF89 is to ascribe
mainly to the adoption of a threshold in the gas density which regulates the process
of star formation, preventing the growth of abundances 
in the outer regions due to the constantly low amount of gas.
We have computed also the evolution of such gradients in time and found that
the inner gradients generally steepen in time.
This is a consequence of the slow formation of the Galactic disk.

\end{itemize}
\bigskip
\noindent
Finally, we would like to discuss briefly the
roles of the different free parameters.
The model presented
here has essentially four free parameters namely $k$, $\nu$,
$\tau_T$ and $\tau_D$. 
The parameters $\nu$ and $k$ of the star formation rate both influence 
the absolute abundances and $k$ influences also the shape of the distribution
of the gas along the disk. 
After performing several numerical experiments,
we found the range of permitted values for
$k$ is 1.5 $\pm$ 0.2. The parameter $\tau_T$ is constrained to be 1 $\pm$ 0.3
Gyr by the [$\alpha$/Fe] plateau and the parameter $\tau_D$ is constrained by
the G-dwarf metallicity distribution. We found that the range of acceptable
values for $\tau_D$ in the solar neighbourhood to be 8 $\pm$ 1.5 Gyrs.

\smallskip\noindent
Our model can be further improved if we take into account the following points:

\begin{itemize}
\item
Preliminary results for the radial profiles
have shown that the SFR and
gas radial distributions along the disk of the Galaxy are flat
 in the outer parts. This behavior is
clearly
a consequence of the star formation threshold. Note, however, that a flat 
gas distribution is observed for $HI$ whereas that of $H_{2}$ 
decreases with the galactocentric distance. As a consequence, we
should consider models with these two gas phases separately (the subject
of a future paper).

\item
The predicted solar and present $^{3}He$ abundance is very high
in comparison with the observed values. This occurs for a large fraction 
of the presently available chemical evolution models (Tosi 1996).
In our models the $^{3}He$ abundance grows
rapidly after the time of the formation of the sun. This is due to the fact
that this is exactly the moment when the stars with masses smaller than $1.4
M_{\odot}$, which are the main producers of $^{3}He$, die.

\item
The predicted helium-to-metals enrichment ratio is 1.6 which
is much smaller than the observed one which
lies between 4 and 5 (Chiappini and Maciel 1994,
Pagel et al. 1992). However, the observed value is 
based on gaseous nebulae abundances which
could be uncertain. 
On the other hand, the predicted value for this
ratio is very dependent on the adopted
nucleosynthesis prescriptions. 
There are many possibilities which could be considered:
a revised $^{4}He$ or heavy element nucleosynthesis, or
the formation of black holes instead of SN explosions in massive stars
above a certain mass.

\item
We should find a more physically justified
parameterization for radial variation 
of the infall timescale in the disk.

\item
We should also investigate the dust influence 
on chemical evolution models (also a subject for a future paper)

\end{itemize}

\acknowledgements
\noindent{\bf acknowledgments} -
C. C. and F. M. want to thank
the SISSA institute and the Department of Astronomy of the
University of Trieste for their kind 
hospitality. 
We also thank the referee Jesper Sommer-Larsen for his useful suggestions.
This work was partially supported by CNPq/Brazil

\vfill\eject
\centerline{\bf Figure Captions}
\smallskip
\noindent
\par\noindent
Fig. 1 - Age metallicity relation for two different data sets.
 The curve shows the best model prediction
\par
\noindent
Fig. 2 - Observed G-dwarf metallicity distributions at the solar vicinity
\par
\noindent
Fig. 3 - [O/Fe] versus [Fe/H] behaviour for model A (solid line) and
 model C (dotted line)
\par
\noindent
Fig. 4 - Temporal evolution of the star formation rate as 
 predicted by model A for the solar vicinity
\par
\noindent
Fig. 5 - Masses ejected in the form of various elements weighted
 by the IMF, are plotted as functions of the initial mass of the progenitor
 star
\par
\noindent
Fig. 6 - Temporal evolution of type Ia and type II SNe as predicted by model
 A for the solar vicinity
\par
\noindent
Fig. 7 - G-dwarf metallicity distribution in the solar vicinity predicted
 by model A. The data are from Rocha-Pinto and Maciel (1995)
\par
\noindent
Fig. 8 - Temporal evolution of the star formation rate, at solar vicinity,
 as predicted by model C
\par\noindent
Fig. 9 - Predicted behaviour of the relative ratios of several heavy
elements to iron with respect to the relative iron abundance, in  the solar 
neighbourhood a) for Mg, b) for Si c) for S and d) for Ca
\par\noindent
Fig. 10 - Same as figure 10 for a) Zn and b) Cu
\par\noindent
Fig. 11 - Same as figure 10 for C
\par\noindent
Fig. 12 - Same as figure 10 for N
\par\noindent
Fig. 13 - log(X/H) versus O/H relation as predicted by model A
\par\noindent
Fig. 14 - [O/Fe] vs [Fe/H] relations for three different galactocentric radii
\par\noindent
Fig. 15 - Radial distribution of the present surface gas density. The data 
are from Rana (1991). Solid line represents model A prediction.

\newpage

\begin{figure}
\figurenum{1a}
\centerline{\psfig{figure=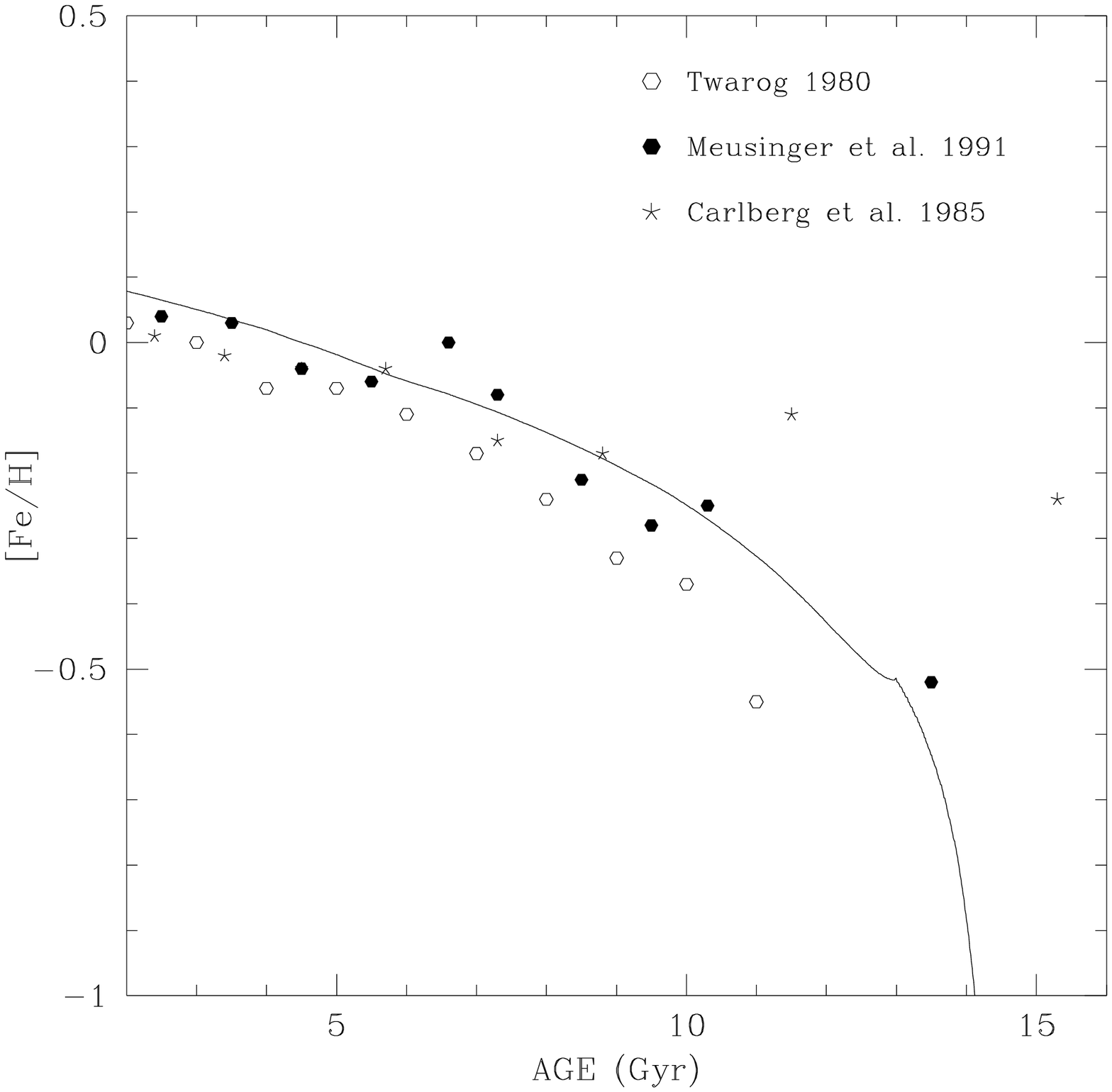,width=14cm,height=19cm} }
\caption{}
\end{figure}

\begin{figure}
\figurenum{1b}
\centerline{\psfig{figure=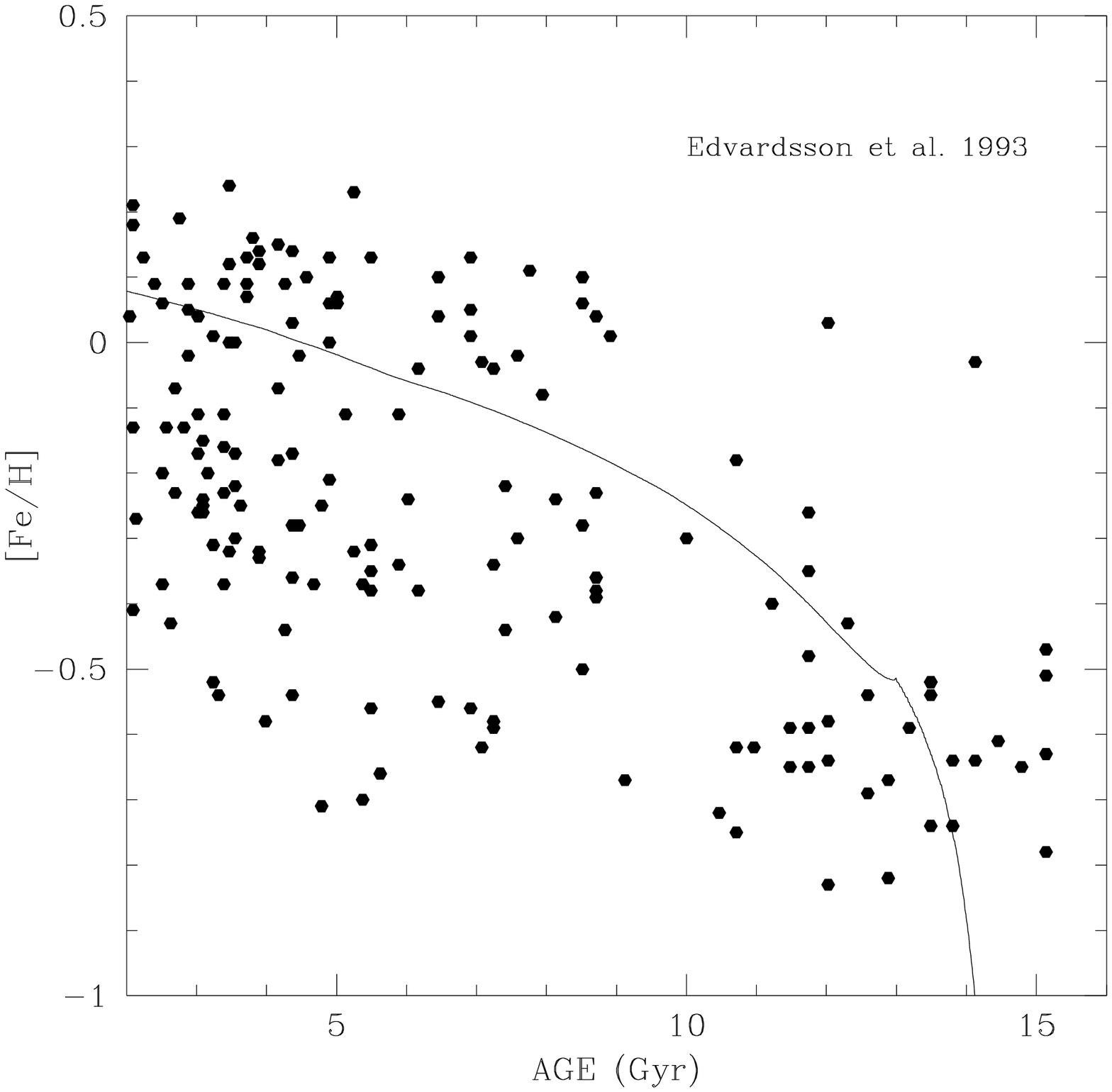,width=14cm,height=19cm} }
\caption{}
\end{figure}

\newpage

\begin{figure}
\figurenum{2}
\centerline{\psfig{figure=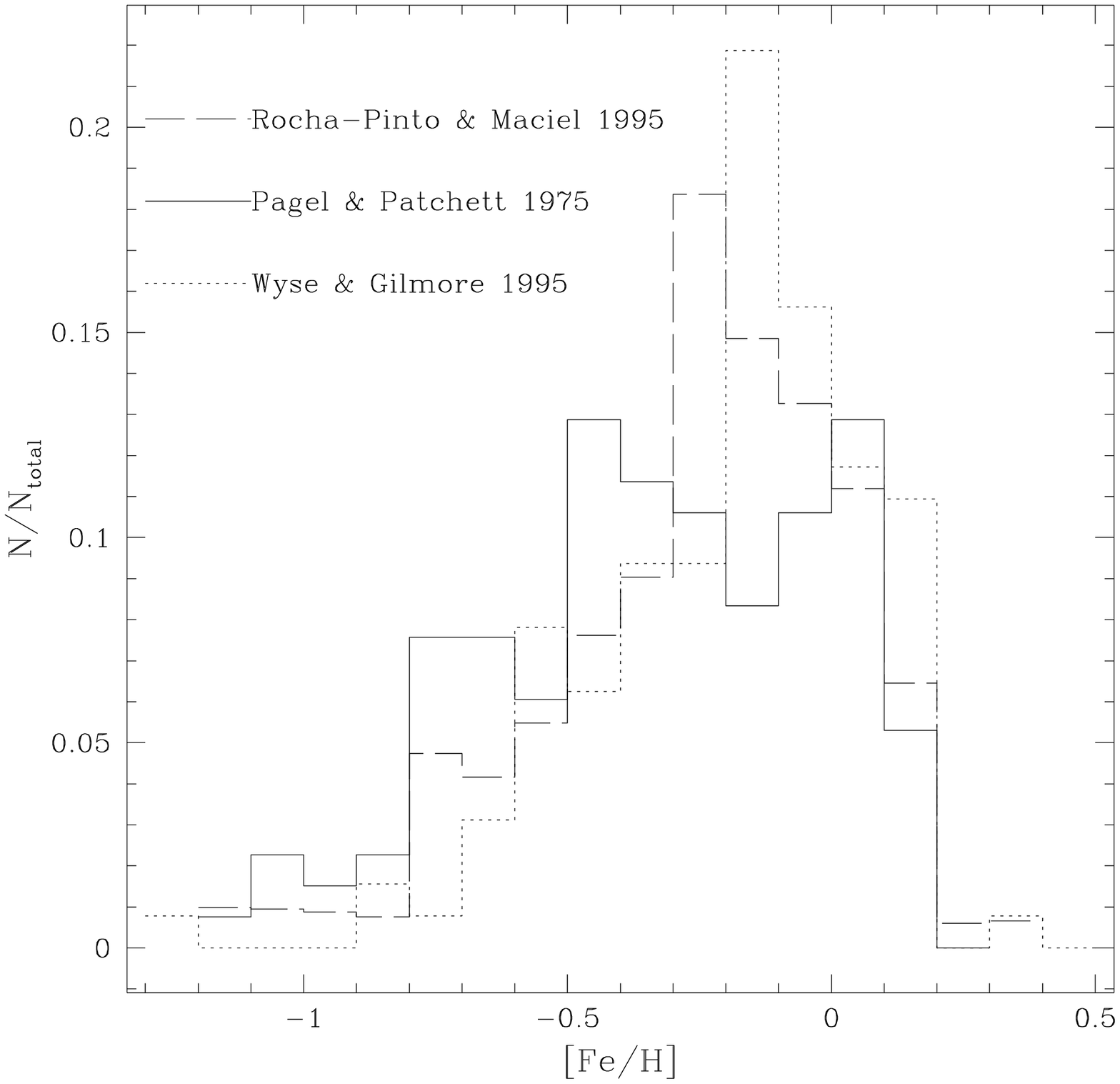,width=14cm,height=19cm} }
\caption{}
\end{figure}

\newpage

\begin{figure}
\figurenum{3}
\centerline{\psfig{figure=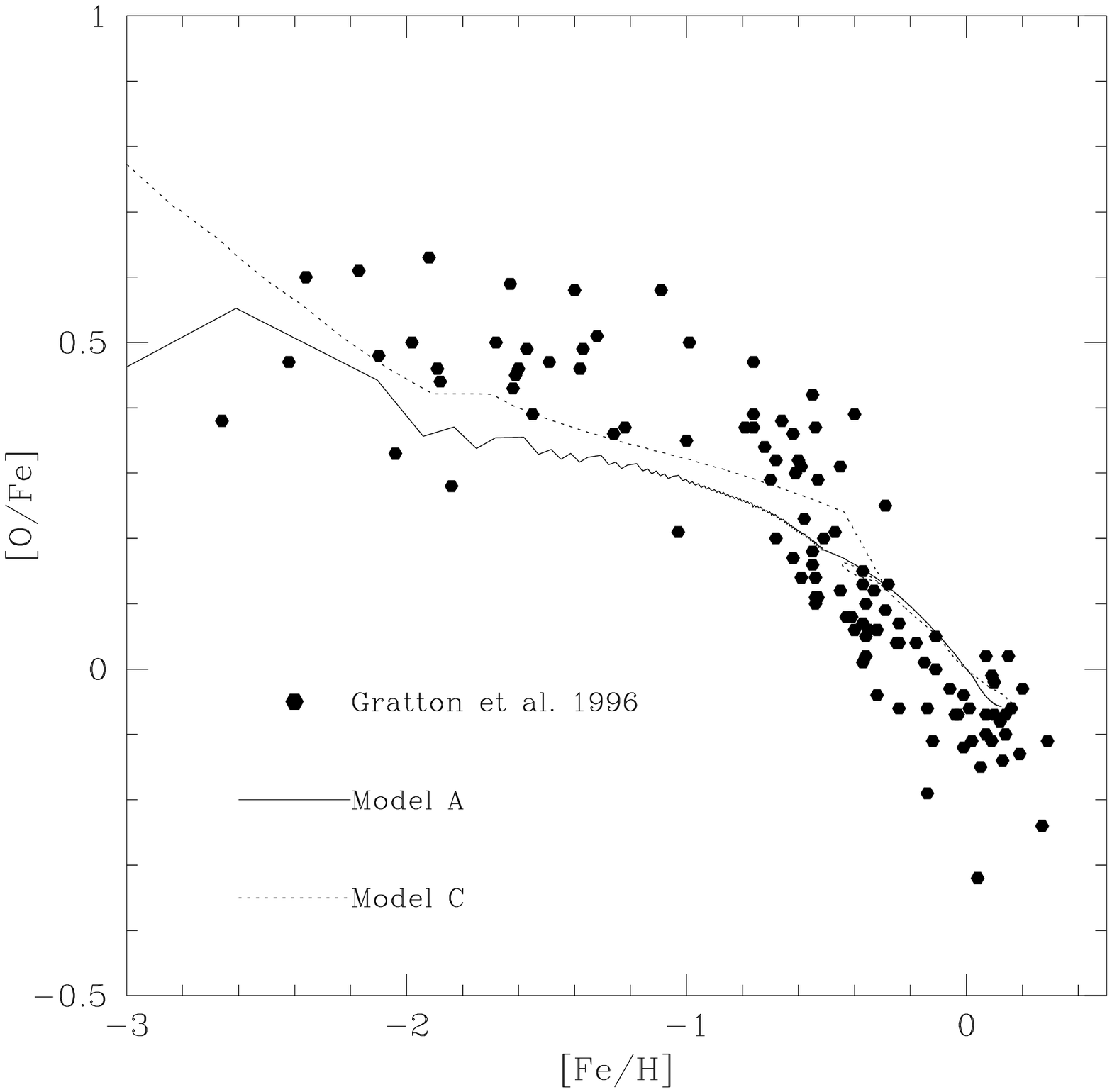,width=14cm,height=19cm} }
\caption{}
\end{figure}

\newpage

\begin{figure}
\figurenum{4}
\centerline{\psfig{figure=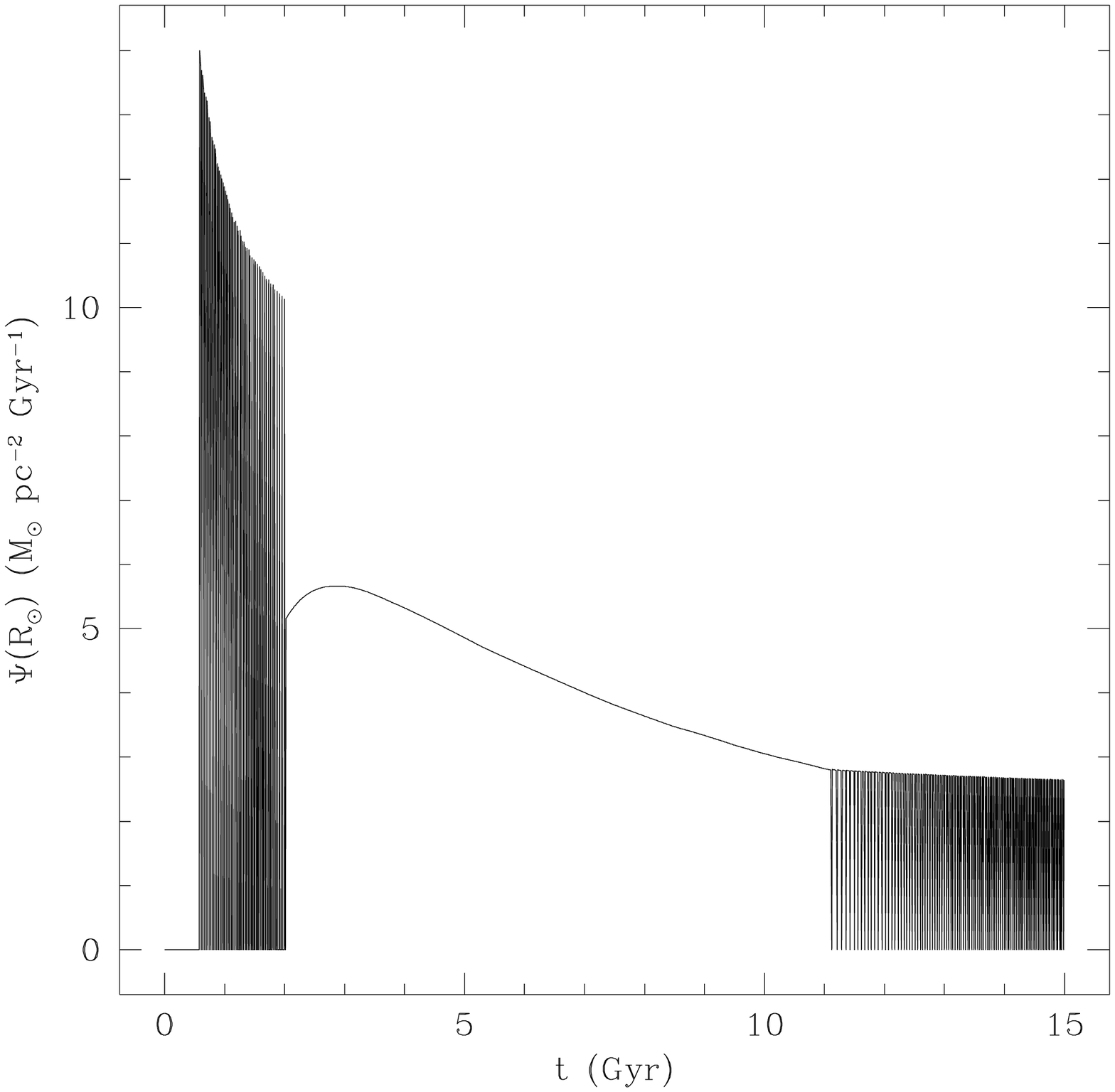,width=14cm,height=19cm} }
\caption{}
\end{figure}

\newpage

\begin{figure}
\figurenum{5}
\centerline{\psfig{figure=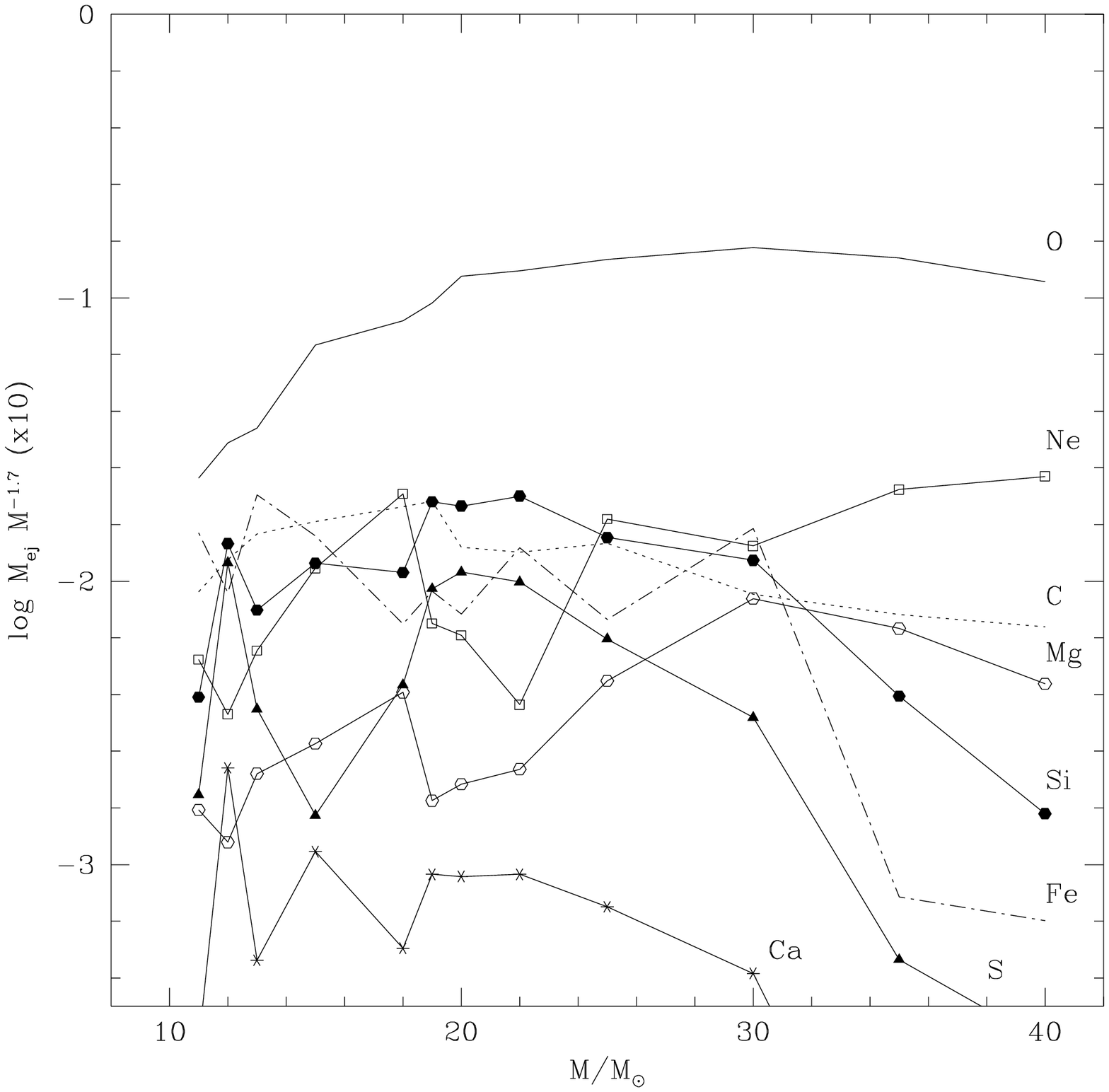,width=14cm,height=19cm} }
\caption{}
\end{figure}

\newpage

\begin{figure}
\figurenum{6}
\centerline{\psfig{figure=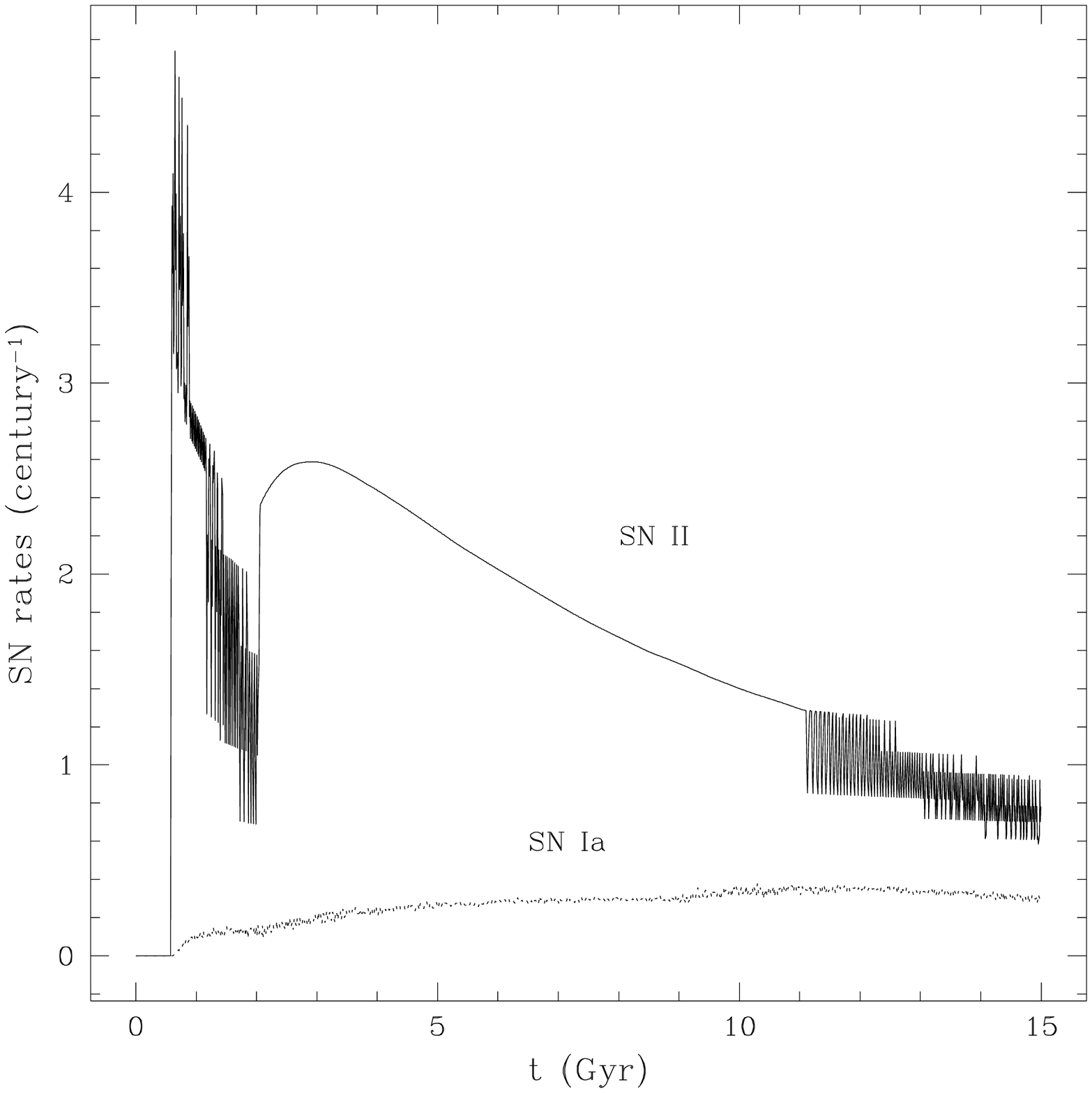,width=14cm,height=19cm} }
\caption{}
\end{figure}

\newpage

\begin{figure}
\figurenum{7}
\centerline{\psfig{figure=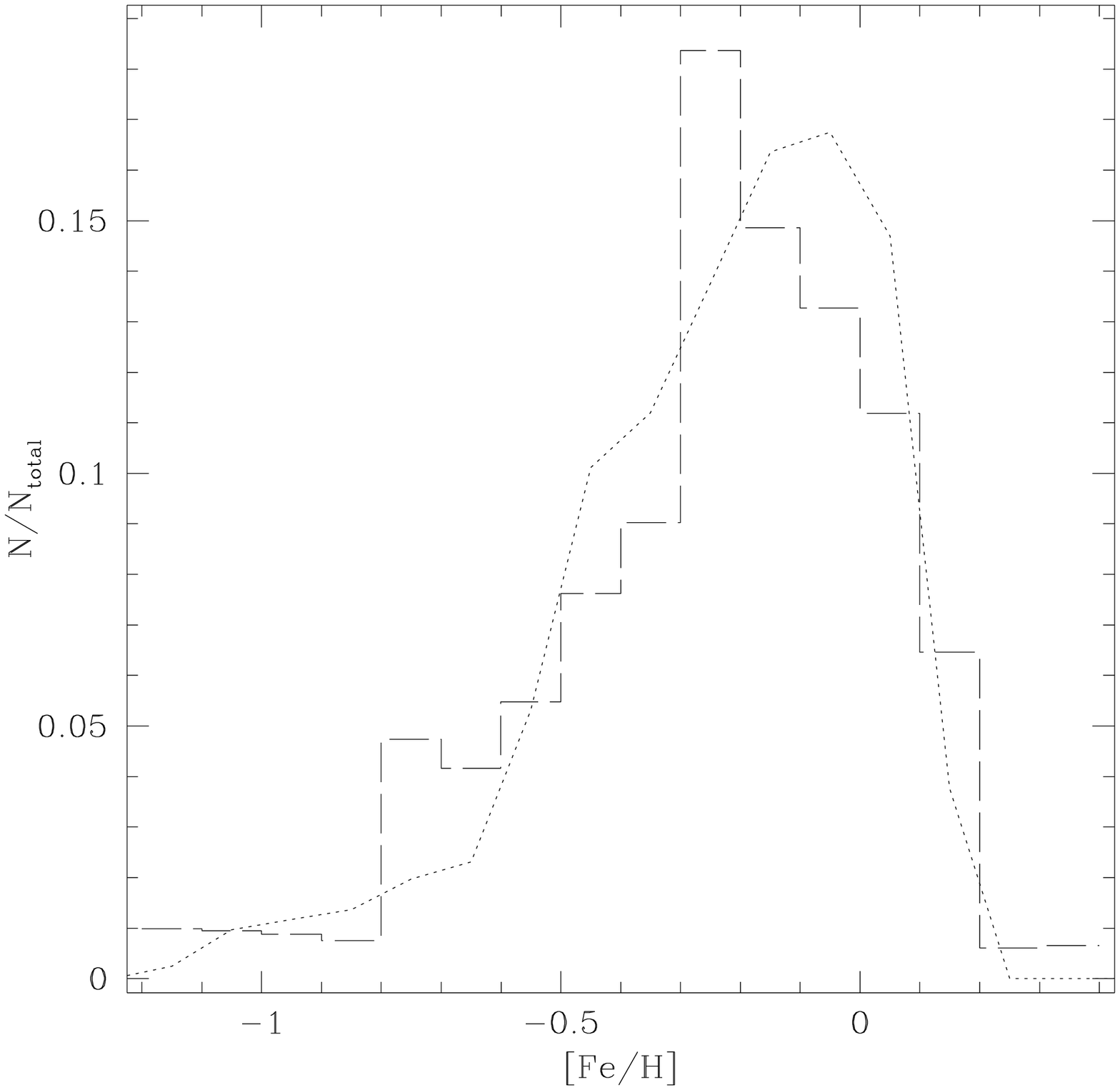,width=14cm,height=19cm} }
\caption{}
\end{figure}

\newpage

\begin{figure}
\figurenum{8}
\centerline{\psfig{figure=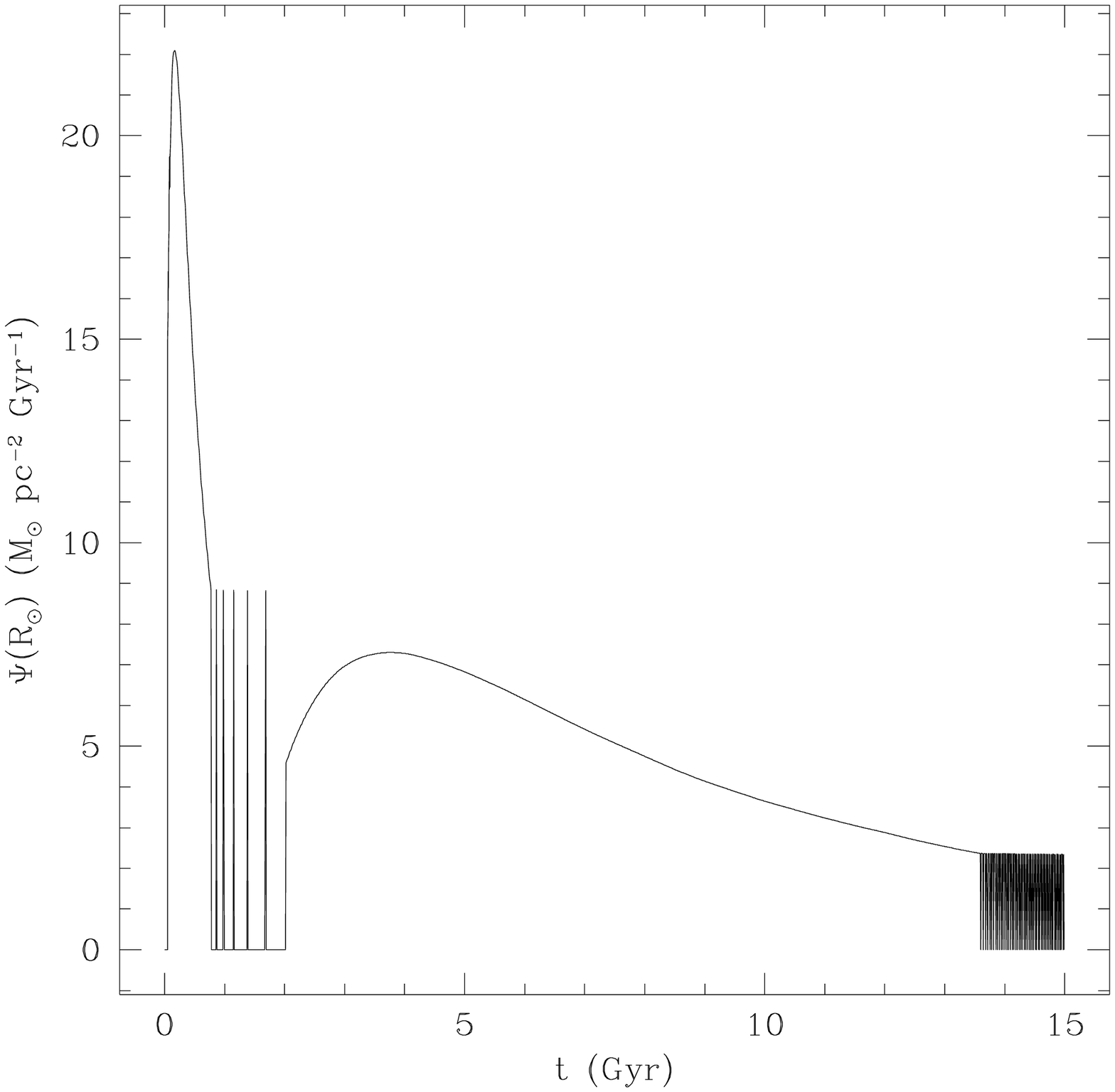,width=14cm,height=19cm} }
\caption{}
\end{figure}

\newpage

\begin{figure}
\figurenum{9a}
\centerline{\psfig{figure=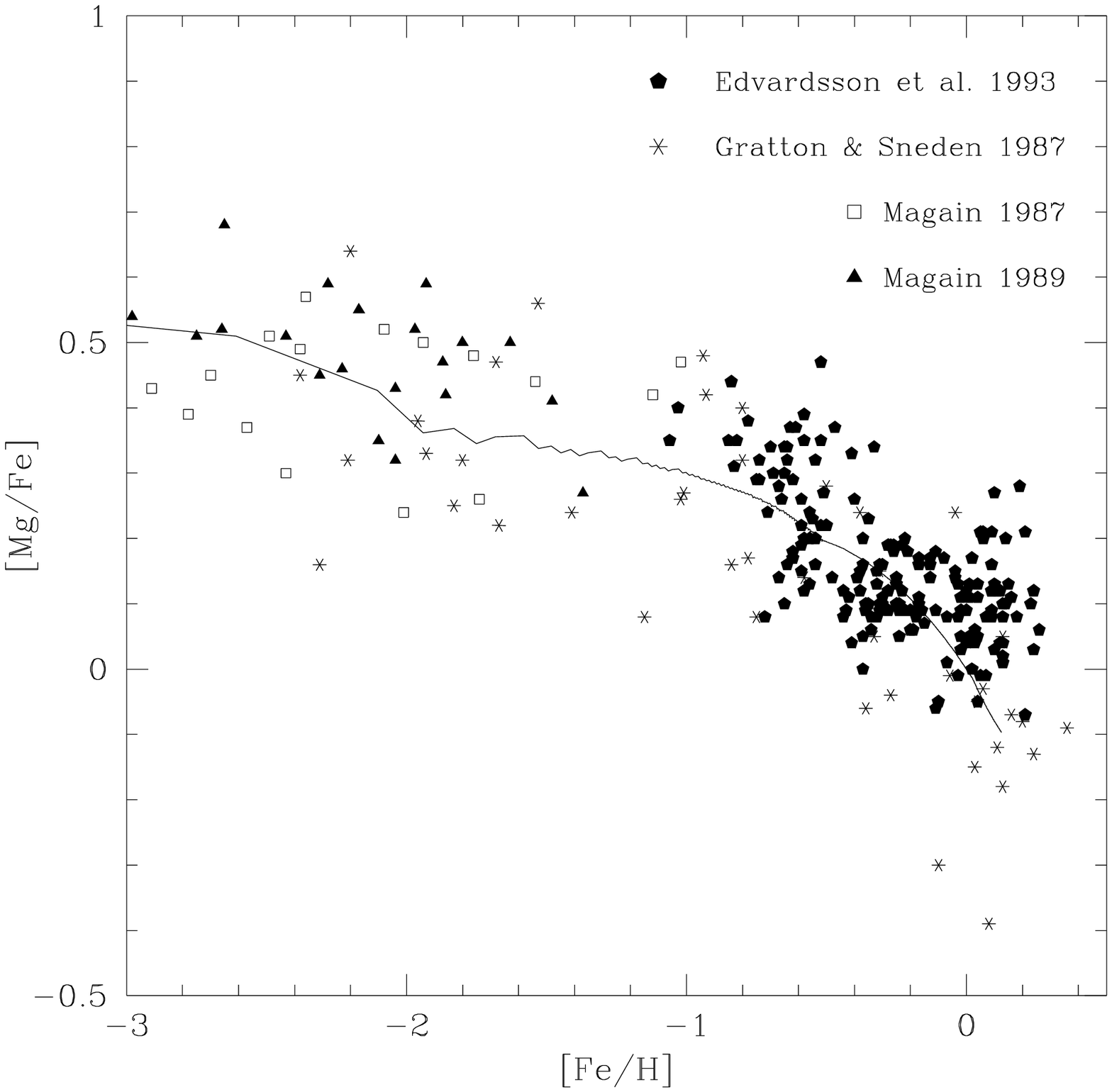,width=14cm,height=19cm} }
\caption{}
\end{figure}

\clearpage

\begin{figure}
\figurenum{9b}
\centerline{\psfig{figure=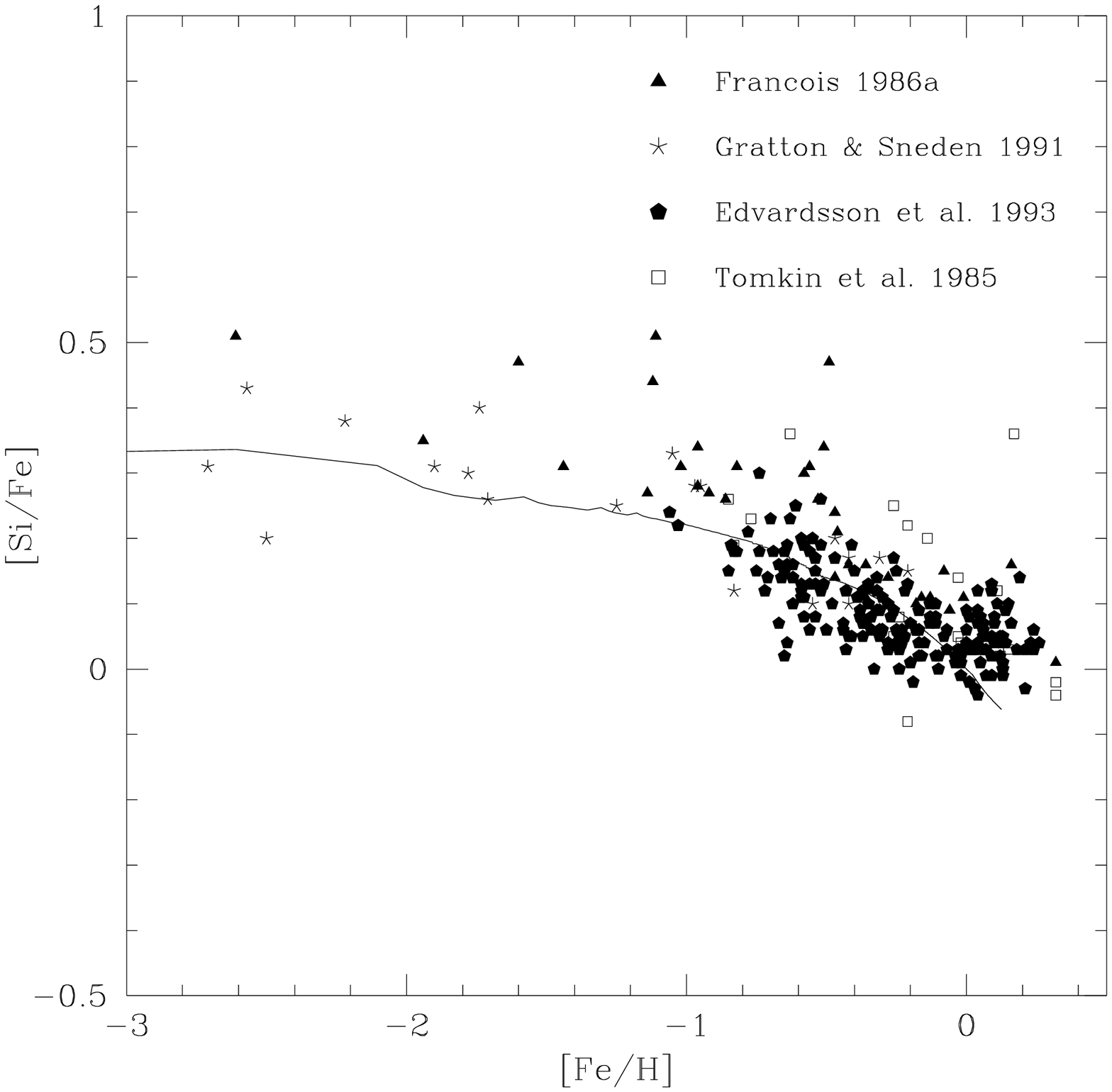,width=14cm,height=19cm} }
\caption{}
\end{figure}

\clearpage

\begin{figure}
\figurenum{9c}
\centerline{\psfig{figure=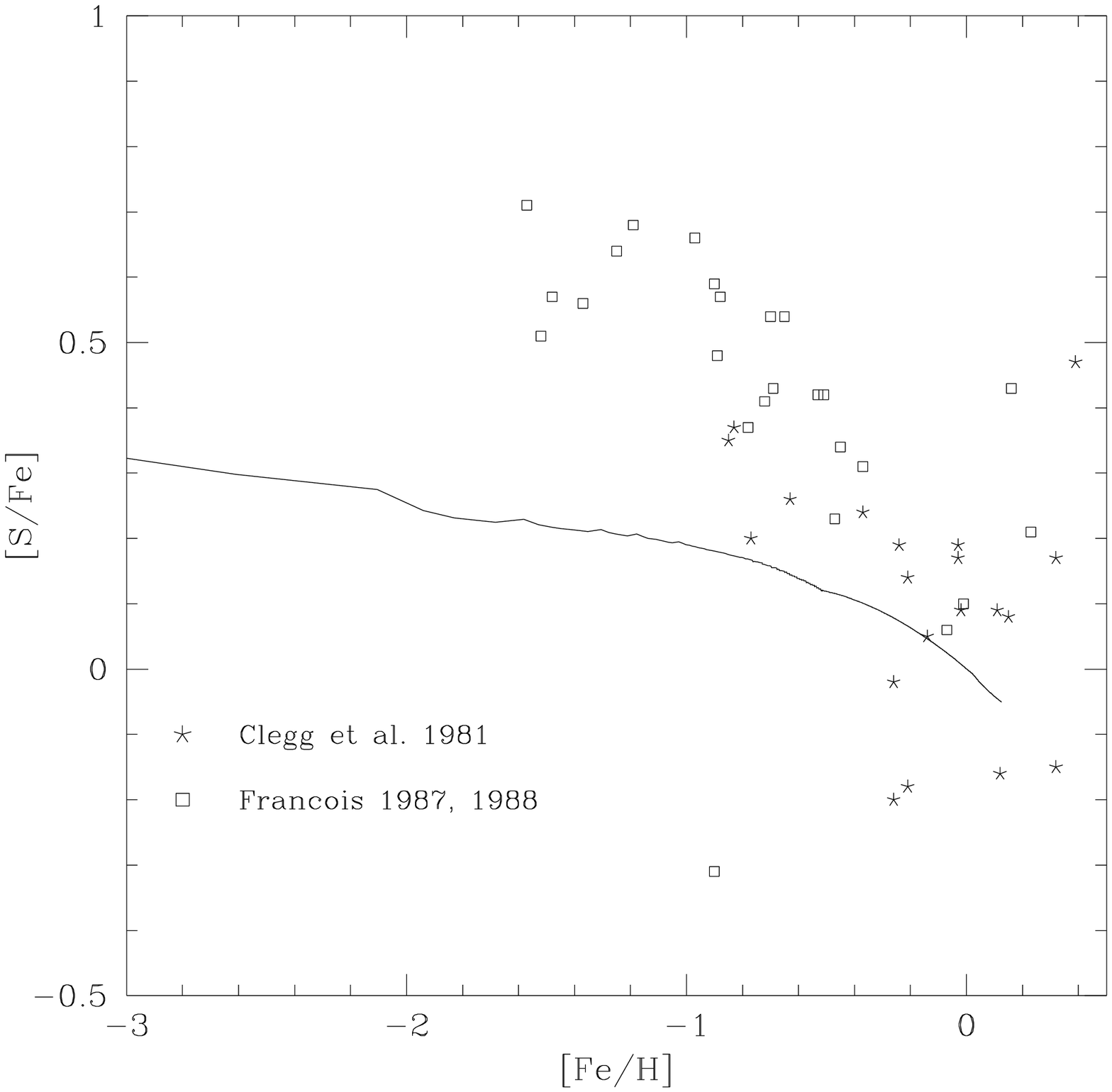,width=14cm,height=19cm} }
\caption{}
\end{figure}

\newpage

\begin{figure}
\figurenum{9d}
\centerline{\psfig{figure=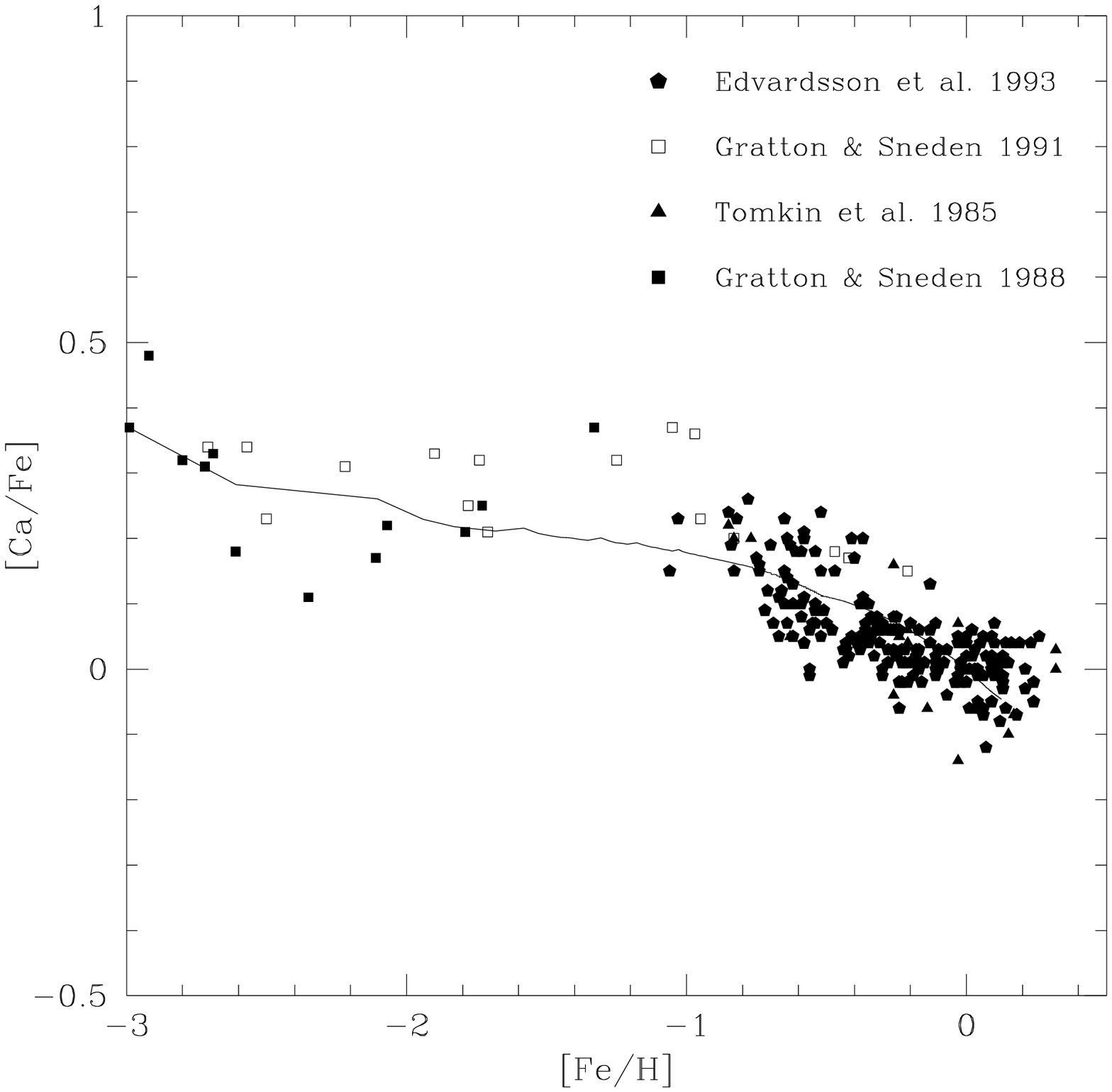,width=14cm,height=19cm} }
\caption{}
\end{figure}

\newpage

\begin{figure}
\figurenum{10a}
\centerline{\psfig{figure=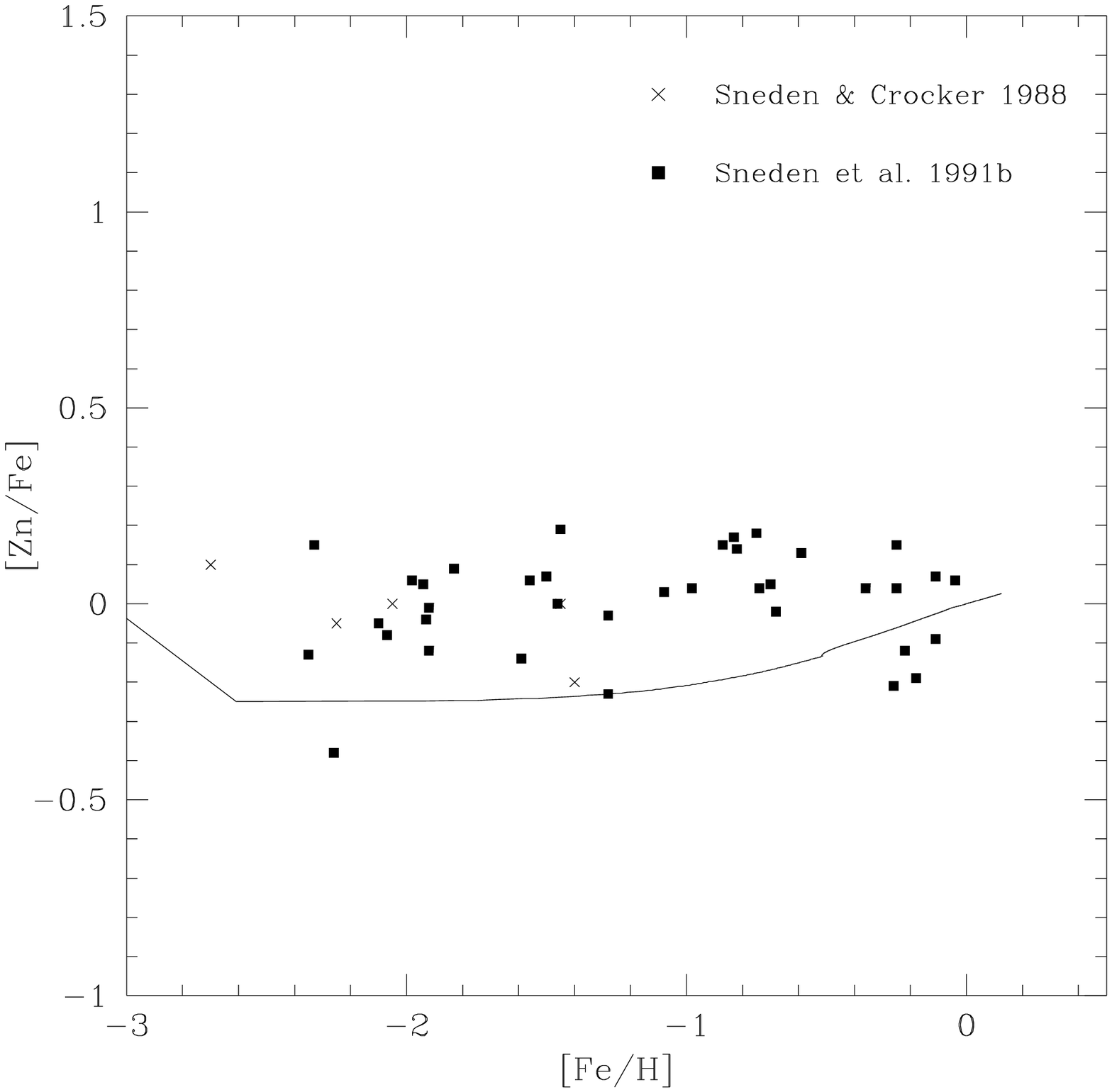,width=14cm,height=19cm} }
\caption{}
\end{figure}

\newpage

\begin{figure}
\figurenum{10b}
\centerline{\psfig{figure=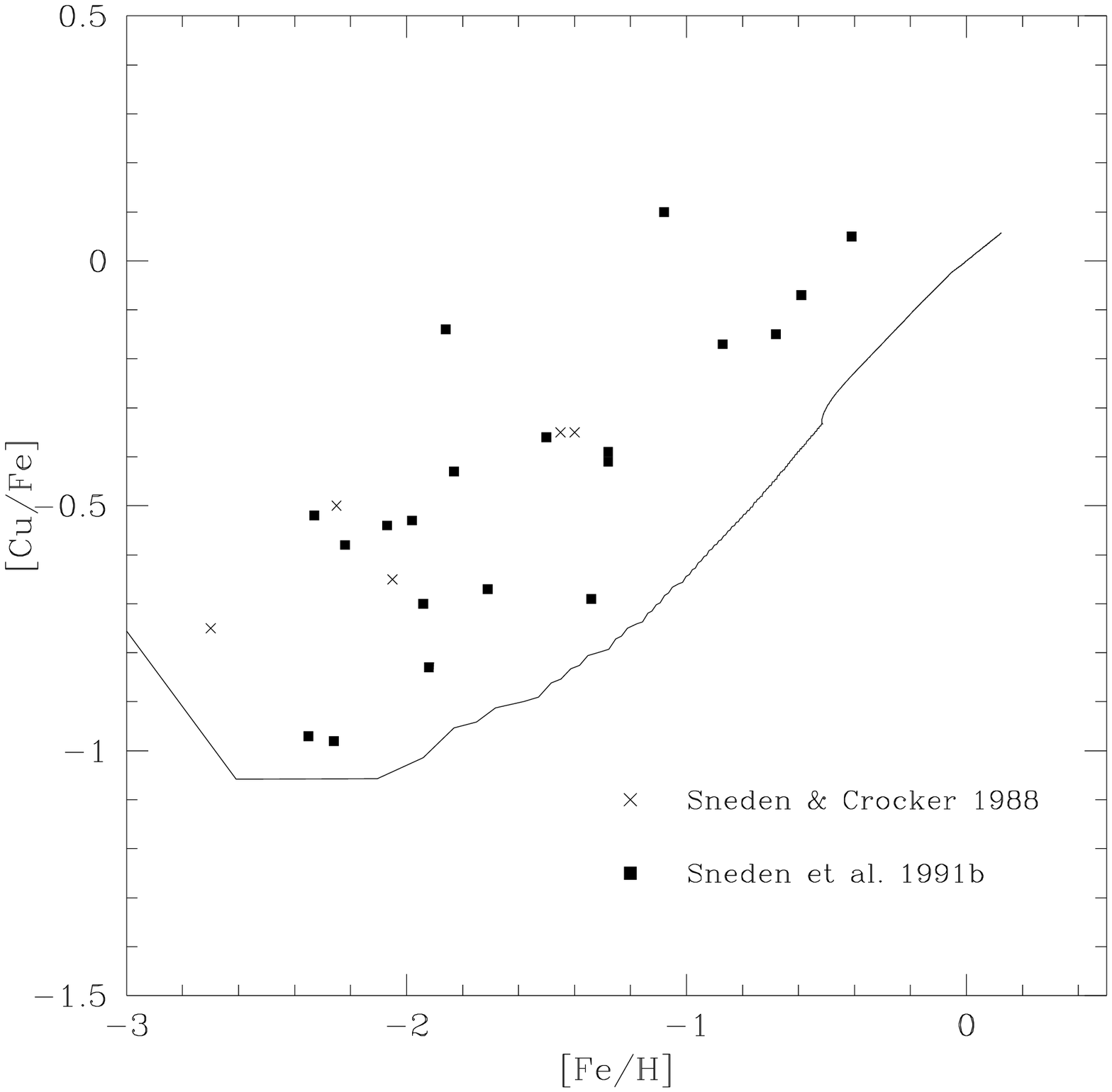,width=14cm,height=19cm} }
\caption{}
\end{figure}

\newpage

\begin{figure}
\figurenum{11}
\centerline{\psfig{figure=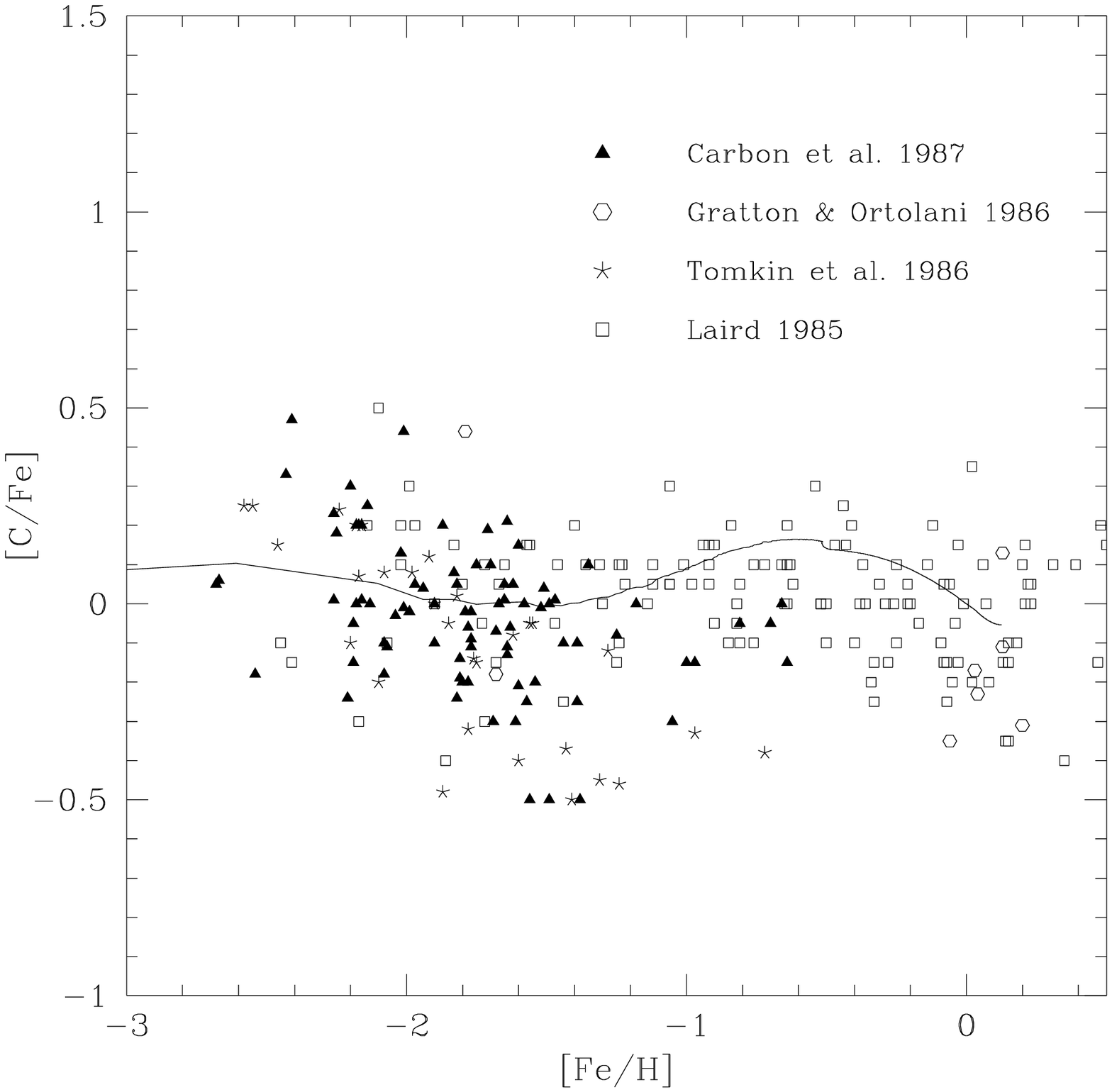,width=14cm,height=19cm} }
\caption{}
\end{figure}

\newpage

\begin{figure}
\figurenum{12}
\centerline{\psfig{figure=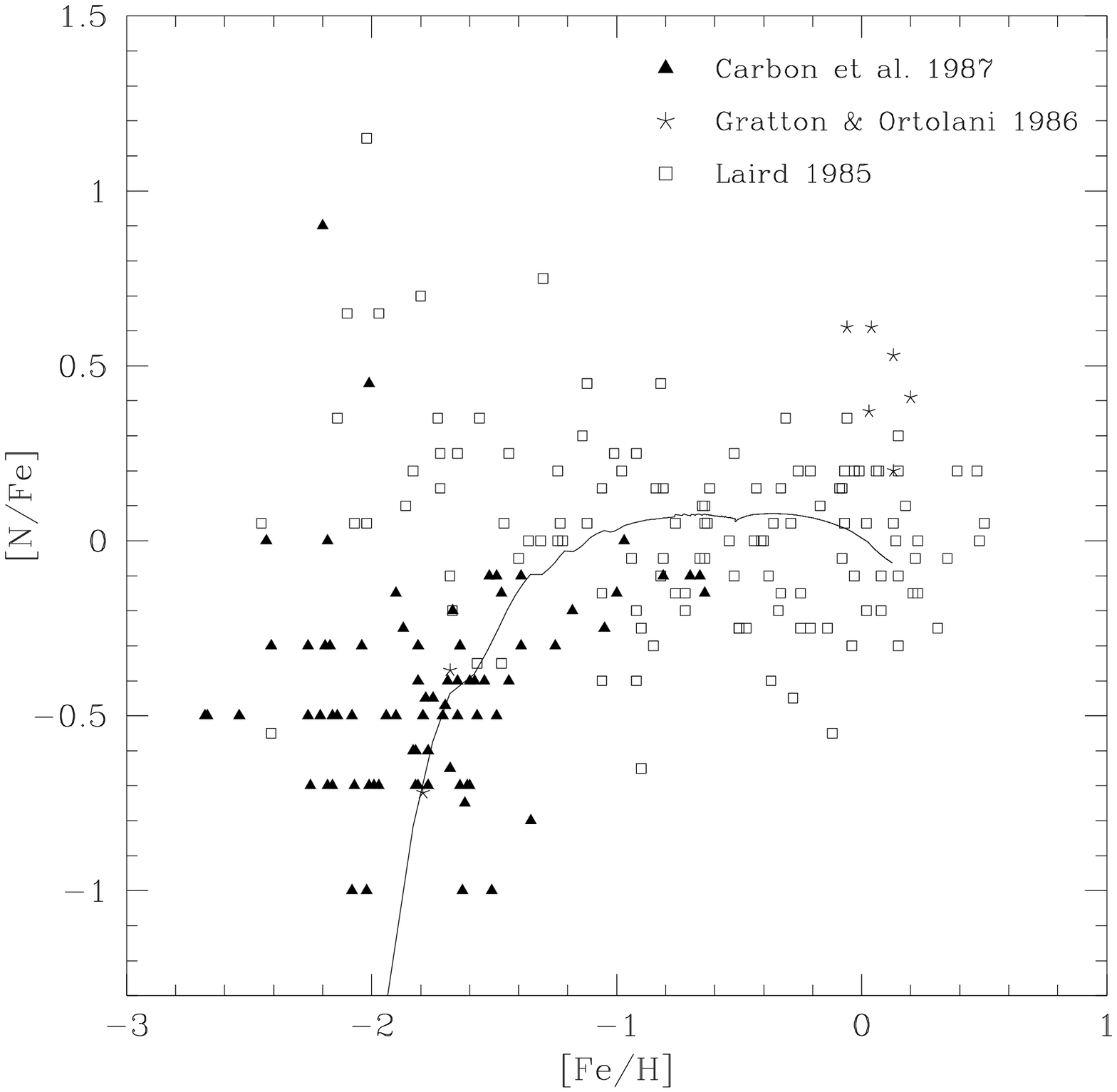,width=14cm,height=19cm} }
\caption{}
\end{figure}

\newpage

\begin{figure}
\figurenum{13}
\centerline{\psfig{figure=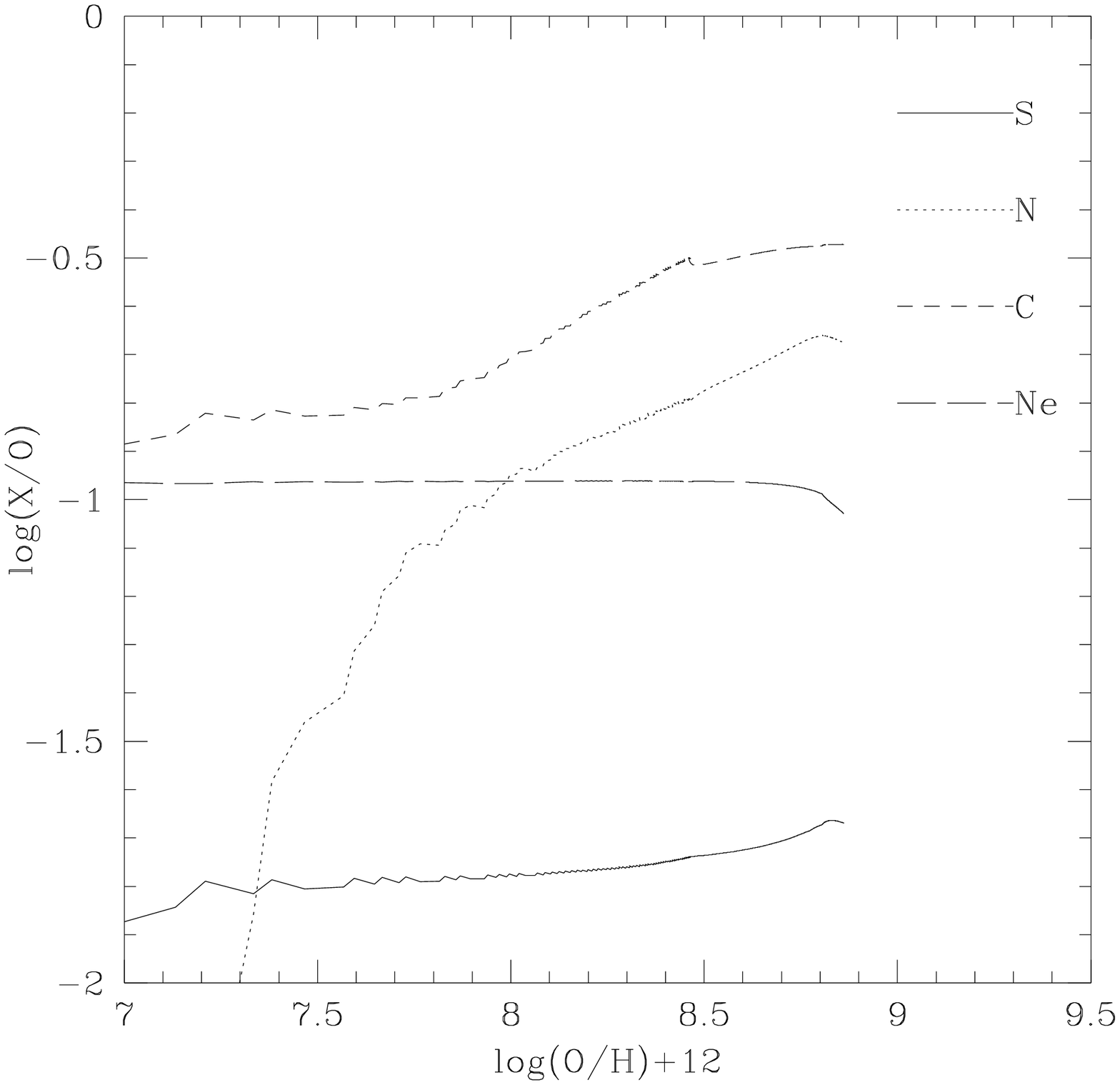,width=14cm,height=19cm} }
\caption{}
\end{figure}

\newpage

\begin{figure}
\figurenum{14}
\centerline{\psfig{figure=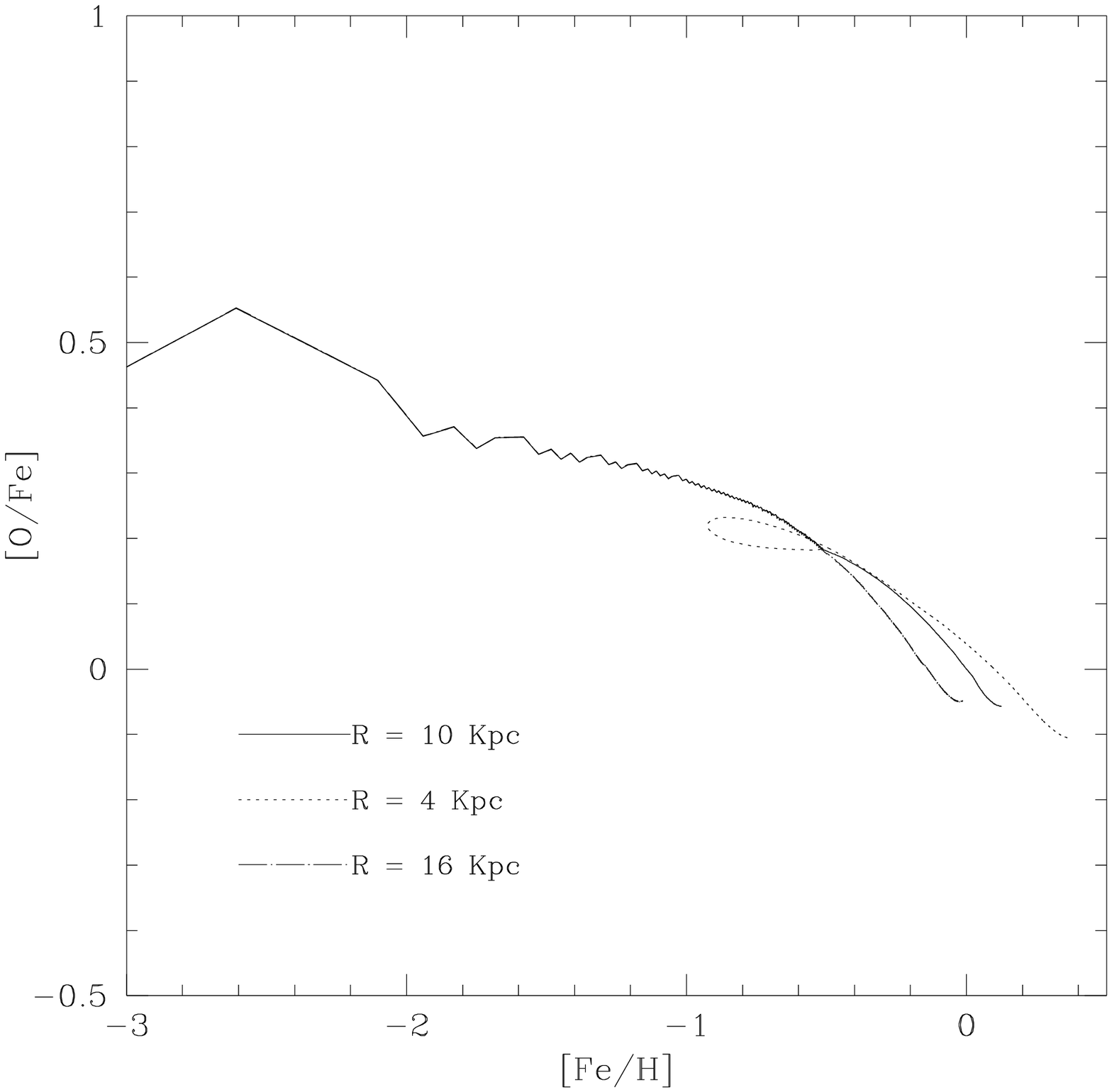,width=14cm,height=19cm} }
\caption{}
\end{figure}

\newpage

\begin{figure}
\figurenum{15}
\centerline{\psfig{figure=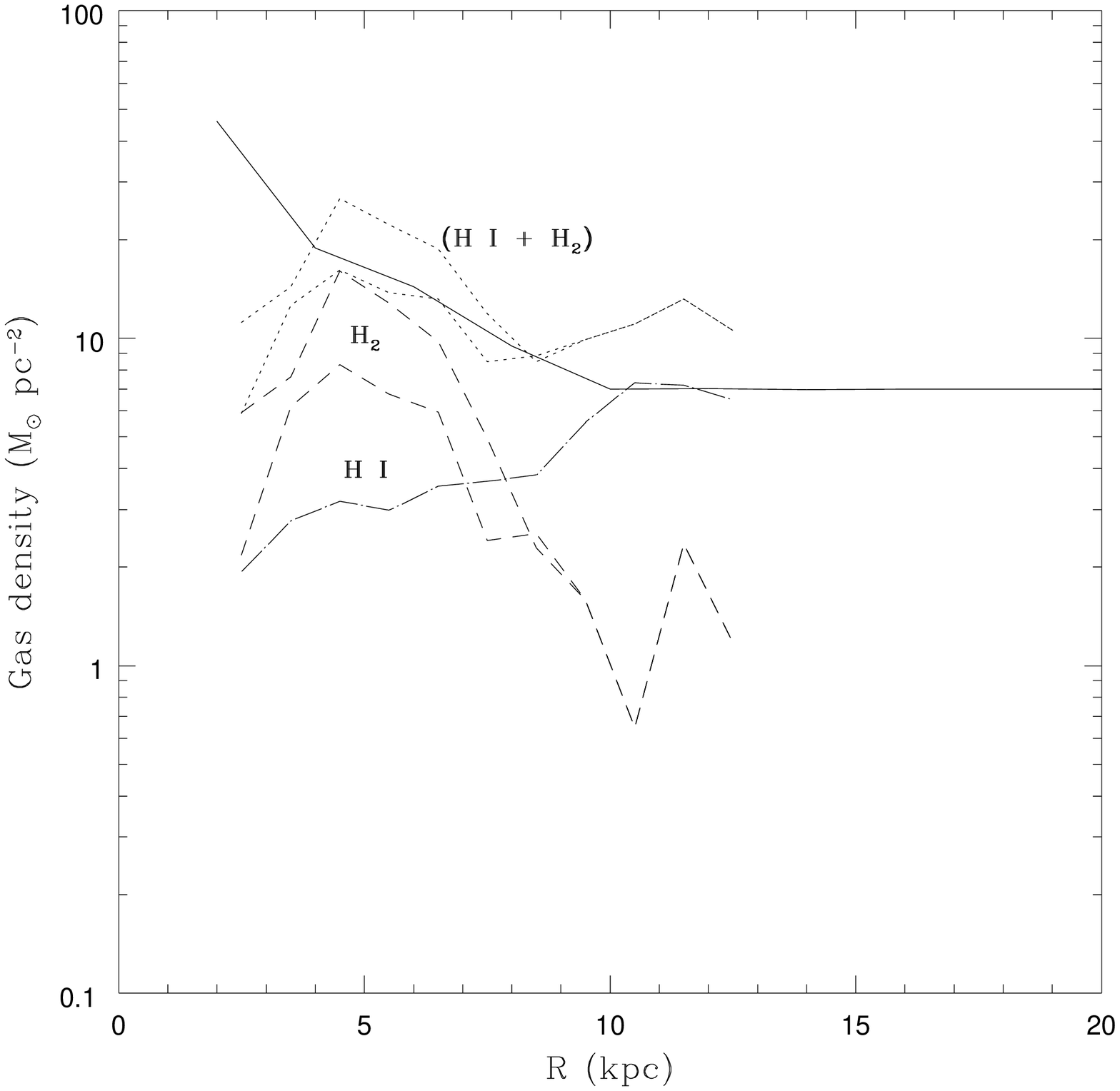,width=14cm,height=19cm} }
\caption{}
\end{figure}

\end{document}